\documentclass[aps,prb,onecolumn,nofootinbib,citeautoscript,10pt]{revtex4-2}

\usepackage{amsmath,amssymb,bm} 
\usepackage{graphicx,comment}

\usepackage[tight]{subfigure} 

\usepackage[dvipsnames]{xcolor} 
\usepackage[papersize={8.5in,11in}]{geometry}
\usepackage[colorlinks=true]{hyperref}
\hypersetup{
    bookmarks=true,         
    unicode=false,          
    pdftoolbar=true,        
    pdfmenubar=true,        
    pdffitwindow=false,     
    pdfstartview={FitH},    
    pdfkeywords={keyword1} {key2} {key3}, 
    pdfnewwindow=true,      
    colorlinks=true,       
    linkcolor=magenta, 
    citecolor=blue,        
    filecolor=magenta,      
    urlcolor=blue           
} 

\usepackage{graphicx}
\usepackage{dcolumn}
\usepackage{color}
\usepackage{amssymb,amsmath}
\usepackage{tabularx,graphicx}
\usepackage{latexsym}
\usepackage{colortbl}
\usepackage{psfrag}
\usepackage{bbm,bm,array,physics}
\usepackage{dsfont}
\usepackage{float, mathrsfs}

\def \nn{\nonumber \\}
\def\*#1{\mathbf{#1}} 

\begin{document}
\title{Signatures of topology in generic transport measurements for Rarita-Schwinger-Weyl semimetals}

\author{Ipsita Mandal$^{1}$}
\email{ipsita.mandal@snu.edu.in}
\author{Shreya Saha$^{1}$}
\author{Rahul Ghosh$^{2}$}

\affiliation{$^{1}$Department of Physics, Shiv Nadar Institution of Eminence (SNIoE), Gautam Buddha Nagar, Uttar Pradesh 201314, India
\\
$^{2}$Department of Physical Sciences, Indian Institute of Science Education and Research Berhampur, Berhampur,
Odisha 760010, India}

\begin{abstract} 
We investigate how the signatures of the topological properties of the bandstructures for nodal-point semimetals are embedded in the response coefficients, arising in two distinct experimental set-ups, by taking the Rarita-Schwinger-Weyl (RSW) semimetal as an example. The first scenario involves the computation of third-rank tensors representing second-order response coefficients, relating the charge/thermal current densities to the combined effects of the gradient of the chemical potential and an external electric field/temperature gradient. On the premises that internode scatterings can be ignored, the relaxation-time approximation leads to a quantized value for the nonvanishing components of each of these nonlinear response tensors, characterizing a single untilted RSW node. Furthermore, the final expressions turn out to be insensitive to the specific values of the chemical potential and the temperature. The second scenario involves computing the magnetoelectric conductivity under the action of collinear electric ($\mathbf E$) and magnetic ($\mathbf B$) fields, representing a planar Hall set-up. In particular, our focus is in bringing out the dependence of the linear-in-$|\mathbf B|$ parts of the conductivity tensor on the intrinsic topological properties of the bandstructure, which are nonvanishing only in the presence of a nonzero tilt in the energy spectrum.
\end{abstract}

\maketitle

\tableofcontents

\section{Introduction}

Three-dimensional (3d) topological semimetals \cite{burkov11_Weyl, yan17_topological, armitage_review, polash-review} form the gapless cousins of the topological insulators (for which a gap-opening implies a topological phase transition), possessing a gapless spectrum. Weyl semimetals (WSMs) exemplify such topological phases, forming an intermediate state in the transition from metals to insulators, in which the conduction and the valence bands touch only at discrete points. This leads to the emergence of zero band-gap and singular points in the Brillouin zone (BZ), known as nodal points, where the density-of-states vanishes exactly. The nodal points are topologically stable, as they cannot be fully gapped out by perturbations that are small (in magnitude) and local in momentum space, such that the bulk gap remains intact sufficiently away from the band-crossing points. The stability of the nodal points and, consequently, the gaplessness of the semimetallic phases are ensured by the fact that these points act as the sources/sinks of topological charges (leading to the notion of defective or singular points). Although the topological charges are determined by the electronic bandstructure of the material, often they are not directly measurable. However, two experimentally-accessible and exploitable properties that appear are:
\begin{enumerate}
\item Quantized response in various transport measurements that depends on the intrinsic Berry curvature (BC) of the bandstructures.
Some examples include circular dichroism \cite{ips-cd1, ips_cd}, circular photogalvanic effect \cite{moore18_optical, guo23_light, kozii, ips_cpge}, and Magnus Hall effect~\cite{papaj_magnus, amit-magnus, ips-magnus}.
\item Fermi-arcs-surface states at the two-dimensional (2d) surface-BZ \cite{armitage_review}.
\end{enumerate}
Although not in the form of a quantized response, the signatures of the BC are very much present in numerous other transport properties, such as intrinsic anomalous Hall effect~\cite{haldane04_berry, goswami13_axionic, burkov14_anomolous}, planar Hall effects \cite{zhang16_linear, chen16_thermoelectric, nandy_2017_chiral, nandy18_Berry, amit_magneto, das20_thermal, das22_nonlinear, pal22a_berry, pal22b_berry, fu22_thermoelectric, araki20_magnetic, mizuta14_contribution, ips-serena, timm, onofre, ips_rahul_ph_strain, rahul-jpcm, ips-kush-review, ips-ruiz, ips-rsw-ph}, magneto-optical conductivity under strong (quantizing) magnetic fields~\cite{gusynin06_magneto, staalhammar20_magneto, yadav23_magneto}, and transmission of quasiparticles across barriers/wells of electric potential  \cite{ips_aritra, ips-sandip, ips-sandip-sajid, krish-sandwich, ips-jns}.

In the WSMs \cite{burkov11_Weyl, yan17_topological}, we encounter twofold band-crossing singularities, with each band having an isotropic linear-in-momentum dispersion in the vicinity of the nodal point. The 230 space-groups in nonrelativistic condensed matter physics allow the possibility of richer bandstructures in the form of symmetry-protected multifold band-crossings \cite{bernevig}. The simplest case is the one where each band exhibits an isotropic linear-in-momentum dispersion, akin to the WSMs. In general, the effective low-energy Hamiltonian of such a system is captured by $\sim \mathbf{k} \cdot \boldsymbol{\mathcal{S}} $, where $\boldsymbol{\mathcal{S}}$ is the vector operator comprising the three components of the angular momentum in the spin-$\varsigma$ representation of the SU(2) group. This stems from the $(2\, \varsigma + 1) $ bands touching at the nodal point, resulting in the itinerant electronic degrees of freedom represented by quasiparticles with pseudospin quantum numbers equalling $ \varsigma  $. The terminology of ``pseudospin'' is used in order to clearly demarcate it as a quantum number distinct from the relativistic spin of an electron. While WSMs feature $\varsigma = 1/2$, the poster child for multifold cases is the Rarita-Schwinger-Weyl (RSW) semimetal~\cite{bernevig, long, igor, igor2, isobe-fu, tang2017_multiple, ips3by2, ips-cd1, ma2021_observation, ips-magnus, ips-jns, ips_jj_rsw, ips-rsw-ph}, having pseudospin-3/2 (i.e., fourfold band-crossings). This nomenclature is inspired by the fact that, in the branch of theoretical high-energy physics, the Rarita-Schwinger (RS) equation describes the field equation of elementary particles with the relativistic spin of 3/2, which naturally arise in supergravity models \cite{weinberg}. Albeit, these higher-spin fermionic particles appear neither in the standard model of particle physics, nor has been detected experimentally. The RSW quasiparticles thus form nonrelativistic analogues of the elusive RS fermions, arising in the context of solid-state systems. 

Let us now elaborate on the origins of the BC and the quantized-response phenomena in widely different experimental settings.
The \textit{topological} properties of a crystalline bandstructure are inferred by treating the BZ as a closed manifold. A 3d nodal-point semimetal is endowed with a nontrivial topology when the nodes demarcate point-like topological defects, mathematically quantified as the locations of the BC monopoles \cite{fuchs-review, polash-review}. The charge of a BC monopole, therefore, is a topological charge, whose
sign assigns a chirality $\chi$ to the node, with $\chi = + 1$ and $\chi = -1$ labelling the so-called
\textit{right-moving} and \textit{left-moving} \textit{chiral} quasiparticles, respectively. Here, we will adopt the widely-used convention of assigning $\chi=1 $ to the bands with negative energy (with respect to the band-crossing point). The physical picture thus represents that a positively-charged (negatively-charged) monopole acts as a source (sink) for the BC flux lines.
Using all the above definitions, we find that the monopole charge of a specific band is computed by employing the familiar Gauss's law, where we integrate the BC flux over a closed 2d surface enclosing the point-defect. If we project on to the space of a pair of bands with the same magnitude of the pseudospin magnetic quantum numbers, we get a two-level system --- the Chern number ($ \mathcal C $) represents a wrapping number of the map from the 2d closed surface (topologically equivalent to $S^2$) to the Bloch sphere ($S^2$), given by the elements of the second homotopy group $\Pi_2(S^2) = \mathbb{Z}$. Thus, the monopole charges are equivalent to the Chern numbers, when interpreted as the wrapping numbers of the defects. The monopole charges always appear in pairs, with each pair having the values $ \pm \, \mathcal C $, thus satisfying the Nielsen-Ninomiya theorem \cite{nielsen81_no}. Our aim in this paper is to unravel some transport coefficients for an RSW node which show a quantized nature, being proportional to $\mathcal C $, thus revealing the nature of the underlying topology. In this context, it is necessary to point out that the four bands at a single RSW node
have Chern numbers $\pm 1$ and $\pm 3$, which sum up to a net monopole charge of magnitude 4 (considering either the two upper or the two lower bands). Therefore, while a WSM-node harbours a Berry monopole charge of magnitude unity, an RSW hosts a net monopole charge of a higher-integer value. In this paper, we will show how these higher values of charges may show up as quantized response in various experimental set-ups. In fact, the signatures of the existence of RSW quasiparticles are reflected by the large values of the topological charges existing in a range of materials, such as $\mathrm{CoSi}$ \cite{tang2017_multiple, takane2019_observation}, $\mathrm{RhSi}$ \cite{sanchez2019_topological}, $\mathrm{AlPt}$ \cite{schroter2019_chiral}, and $\mathrm{PdBiSe}$ \cite{lv2019_observation}.

In this paper, we consider the quantization associated with two distinct experimental set-ups:
\begin{enumerate}
\item  Nonlinear transport under the effect of an external electric field (or temperature gradient) and the gradient of the chemical potential, constituting a response of electrochemical (or thermochemical) nature \cite{ruiz_electro}. Here, we dub the associated electrical and thermal conductivity tensors as the electrochemical response (ECR) and thermochemical response (TCR), respectively. For a single untilted node of WSM, the ECR has been shown to take quantized values for temperature $T$ equalling zero \cite{ruiz_electro}. However, there were various shortcomings in their steps to derive the form of the response, which we will clarify in the course of our computations. Moreover, we will show the quantized nature of both the ECR and the TCR considering a
\textit{generic temperature}\footnote{Albeit with the constraint $T \ll \mu $ (while using natural units), so that we are in the low-energy limit where the effective continuum model for the node is valid. Since $\mu $ ranges around 0.01-0.1 eV \cite{tang2017_multiple}, we must have $T \lesssim 10^{-3} $ eV. For example, recent experiments \cite{claudia-multifold} probe response setting the temperature ranges of the order of $8.617 \times 10^{-5}$ -- $8.617 \times 10^{-3}$ eV, which are equivalent to $\sim 1 $ -- $100 $ K (in SI units).}
for the fourfold nodal point of an RSW semimetal, which is a multiband generalization of the WSM case.

\item Components of the magnetoelectric conductivity under the action of electric ($\mathbf E $) and magnetic ($\mathbf B $) fields, applied parallel to each other, constituting a planar Hall set-up. Here, we will demonstrate the quantized nature of the linear-in-$B$ parts of the linear-response tensor, arising from the four bands of an RSW node.
 \end{enumerate}

\begin{figure}[]
\centering
\subfigure[]{\includegraphics[width=0.2 \textwidth]{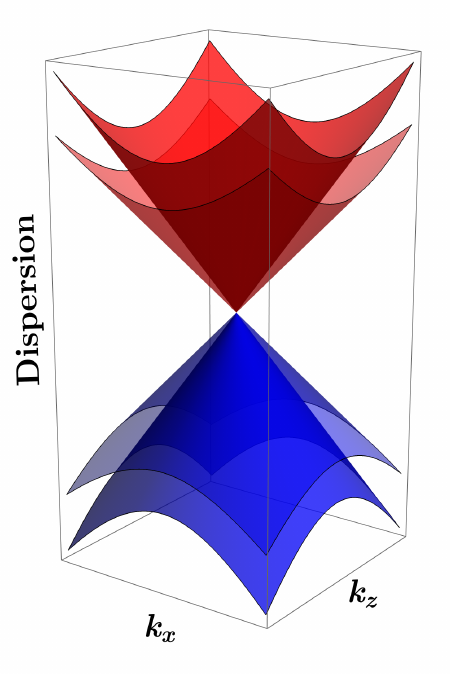}} \hspace{ 2 cm}
\subfigure[]{\includegraphics[width=0.2 \textwidth]{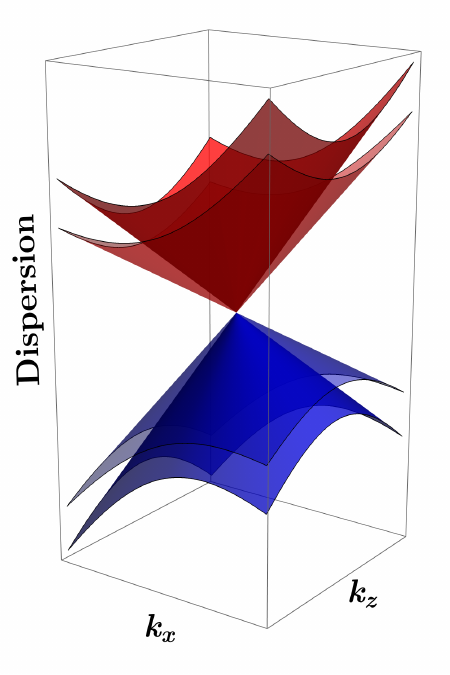}}
\caption{\label{figdisp}
Schematics of the dispersions of the four bands at a single RSW node: Subfigures (a) and (b) represent an untilted node (i.e., $\eta = 0$) and a tilted node (with $ 0<|\eta |< 1$), respectively.
}
\end{figure}

The paper is organized as follows: In Sec.~\ref{secmodel}, we outline the form of the low-energy effective Hamiltonian in the vicinity of an RSW node and, also, show the expressions for various topological quantities. While Sec.~\ref{secnonlinear} deals with the nonlinear response coefficients dubbed as ECR and TCR, Sec.~\ref{sechall} focusses on the linear response associated with the magnetoelectric conductivity. In the end, we wrap up with a summary and outlook in Sec.~\ref{sec_summary}. In Appendix~\ref{secboltz}, the details of the derivations of the response tensors, applying the semiclassical Boltzmann formalism, are provided. The remaining appendices are devoted to elaborating on some of the details of the intermediate steps, necessary to derive the final expressions shown in the main text. In all our expressions, we resort to using the natural units, which implies that the reduced Planck's constant ($\hbar $), the speed of light ($c$), and the Boltzmann constant ($k_B $) are each set to unity.

\section{Model} 
\label{secmodel}

With the help of group-theoretic symmetry analysis and first-principles calculations, it has been shown that seven space-groups may host fourfold band-crossing points \cite{bernevig} at high-symmetry points of the BZ. Nearly 40 candidate materials have also been identified that can host the resulting RSW quasiparticles.
Ref.~\cite{grushin-multifold} has tabulated the multifold degneracies in the 65 chiral space-groups, characterizing the chiral crysals, which are the ones with only orientation-preserving symmetries. A chiral fourfold band-crossing can be realized in the space-groups (1) 195–198 and 207–214 at the $\Gamma $-point; (2) 207 and 208 at the $R$-point; and (3) 211 and 214 at the $H$-point, in the presence of spin-orbit coupling. These fourfold degeneracies exhibit a BC texture that is homotopic to that of a spin-$3/2$ moment in a magnetic field. For an RSW node, the effective Hamiltonian possesses a full SU(2) invariance (i.e., a full rotational invariance). It has been shown \cite{prb108035428} that chiral topological metals belonging to the SrGePt family (e.g., SrSiPd, BaSiPd, CaSiPt, SrSiPt, BaSiPt, and BaGePt), characterized by the space-group 198, host RSW quasiparticles, sixfold excitations (two copies of pseudospin-1 fermions), as well as Weyl points in their bandstructures, when spin-orbit coupling is considered. More explicitly, a fourfold-degenerate node appears at the center of the BZ (i.e., the $\Gamma $-point), carrying the monopole charge of $+ \, 4 $, while a sixfold-degenerate node arises at the boundary of the BZ (i.e., the $R$-point) with a net monopole charge equalling $- \,2-2 = - \, 4$. Here, we ignore any internode scatterings, which is justified because the energy offset between the fourfold-degenerate point (at $\Gamma$) and the sixfold-degenerate point (at $R$) is usually large \cite{tang2017_multiple, prb108035428, prl119206401, yamakage}.

The usual method of linearizing the $\mathbf{k} \cdot \mathbf {p}$ Hamiltonian about such a degeneracy point provides us with the low-energy effective continuum Hamiltonian, valid in the vicinity of the node. The explicit form of this Hamiltonian, for a single node with chirality $\chi $, is given by 
\begin{align}
\mathcal{H}(\mathbf{k})= \eta \, v_0 \,  k_z 
+ \chi \, v_0 \left( 
k_x\,	{\mathcal J }_x + k_y\, {\mathcal J }_y
+  k_z \, {\mathcal J }_z \right),
\end{align}
where $ \boldsymbol{\mathcal J } = \lbrace {\mathcal J }_x,\, {\mathcal J }_y,\, {\mathcal J }_z \rbrace $ represents the vector operator whose three components comprise the angular momentum operators in the spin-$3/2$ representation of the SU(2) group. Here, $\eta $ represents the tilt parameter, with $|\eta |< 1$ representing the type-I phase, which is the scenario under consideration. We choose the commonly-used representation where
\begin{align}
{\mathcal J }_x= 
\begin{pmatrix}
	0 & \frac{\sqrt{3}}{2} & 0 & 0 \\
	\frac{\sqrt{3}}{2} & 0 & 1 & 0 \\
	0 & 1 & 0 & \frac{\sqrt{3}}{2} \\
	0 & 0 & \frac{\sqrt{3}}{2} & 0 
\end{pmatrix} , \quad
{\mathcal J }_y=
\begin{pmatrix}
	0 & \frac{-i \,  \sqrt{3}}{2}  & 0 & 0 \\
	\frac{i \, \sqrt{3}}{2} & 0 & -i & 0 \\
	0 & i & 0 & \frac{-i \, \sqrt{3}}{2}  \\
	0 & 0 & \frac{i \, \sqrt{3}}{2} & 0 
\end{pmatrix}, \quad
{\mathcal J }_z =
\begin{pmatrix}
	\frac{3}{2} & 0 & 0 & 0 \\
	0 & \frac{1}{2} & 0 & 0 \\
	0 & 0 & -\frac{1}{2} & 0 \\
	0 & 0 & 0 & -\frac{3}{2} 
\end{pmatrix}.
\end{align}
Our convention is such that the tilting is with respect to the $k_z$-direction.
The energy eigenvalues are found to be
\begin{align}
\label{eqeval}
\varepsilon_s ( k ) =  s \, v_0 \, k + v_0 \, \eta \, k_z \,, \quad 
s \in  \left \lbrace  \pm \frac{1}{2}, \pm \frac{3}{2} \right \rbrace,
\end{align}
where $ k = \sqrt{k_x^2 + k_y^2 + k_z^2 } $.
Hence, each of the four bands has a linear-in-momentum dispersion (cf. Fig.~\ref{figdisp}). The signs of ``$+$'' and ``$-$'' give us the dispersion relations for the conduction and valence bands, respectively. The corresponding group velocities of the quasiparticles are given by 
\begin{align} 
\boldsymbol{v}_{s}(\mathbf{k}) &= 
\nabla_{\mathbf{k}}  \varepsilon_{s}(\mathbf{k})  
=  \frac{ s \, v_0 \,\mathbf k }{k} 
+ \eta \, v_0 \, {\mathbf{\hat z}}  \, .  
\end{align}

\subsection{Properties with topological origins}

A nontrivial topology of the bandstructure of the RSW semimetals gives rise to the BC and the orbital magnetic moment (OMM), using the starting expressions of \cite{xiao_review}
\begin{align} 
	{\mathbf \Omega} _{s}( \mathbf k) &=  i \,
	 \langle \nabla_{ \mathbf{k}} \psi _{s} ({ \mathbf{k}})| \, \cross \, 
	 | \nabla_{ \mathbf{k}} \psi _{s} ({ \mathbf{k }}) \rangle 
\text{ and } 
	\mathbf{m} _{s}( \mathbf k) =
- \,\frac{e}{2} \, \text{Im} \left[ \langle \nabla_\mathbf{k} 
\psi _{s} \vert  \cross \left( \mathcal{H} {(\mathbf{k})} 
 - \varepsilon_{s}(\mathbf{k}) \right) 
	\vert  \nabla_\mathbf{k} \psi _{s} \rangle \right],
\end{align}  
respectively. Here, $| \psi _{s} ({ \mathbf{k }}) \rangle $ denotes the eigenfunction for the $s^{\rm th}$ band at the node with chirality $\chi $, and $e$ denotes the magnitude of the charge of a single electron. 
Evaluating these expressions for the RSW node described by $\mathcal{H}(\mathbf{k}) $, we get
\begin{align}  
\label{eqberry}
	{\mathbf \Omega} _{s}( \mathbf k) &=    
	 - \, \frac{\chi \,   s \,\mathbf k  } {k^3}  \text{ and } 
\mathbf{m} _{s}( \mathbf k) 
= - \, \frac{e \, \chi\, v_0 \, \mathcal{G}_s  \,\mathbf k } {k^2} \, ,
\text{ with } \mathcal{G}_{\pm\frac{3}{2}} = \frac{3}{4} \text{ and } 
\mathcal{G}_{\pm\frac{1}{2}} = \frac{7}{4} \,.
\end{align}
Since $\mathbf  \Omega _{s}( \mathbf k)$ and $\mathbf{m} _{s}( \mathbf k)$ are the intrinsic properties of the bandstructure, they depend only on the wavefunctions. Clearly, they are related as
\begin{align}
	\mathbf{m} _{s}( \mathbf k) = 
	\frac{e \, v_0 \, \mathcal{G}_s \, k } {s} 
	\, {\mathbf \Omega} _{s}( \mathbf k) \,.
\end{align}
We observe that, unlike the BC, the OMM does not depend on the sign of the energy of the dispersion. 
Since the tilt parameter enters the Hamiltonian as the coefficient of the $4 \cross 4$ identity matrix, its presence or absence does not affect the eigenspinors and, hence, the topology of the low-energy
theory in the vicinity of the band-crossing point. Clearly, the BC or OMM of an RSW node does not depend
on the tilt parameter.

The OMM behaves exactly like the electron spin because, on applying a magnetic field $\mathbf B$, it couples to the field through a Zeeman-like term, quantified by 
\begin{align} 
\label{eqomme}
\zeta _{s} ({\mathbf k})  &= - \, \mathbf{m} _{s}  (\mathbf k) \cdot \mathbf{B}
=  e \, \chi\, v_0 \, \mathcal{G}_s  \,   
\frac{  \mathbf{k}   \cdot  \mathbf{B} } {k^2} \, .
\end{align}
Therefore, we have 
\begin{align} 
\label{eqtoten}
\xi_s ({\mathbf k}) & = \varepsilon_{s}(\mathbf{k})  + \zeta _{s}(\mathbf{k}) 
\,, \quad 
	\boldsymbol{w}_s(\mathbf{k})= 
	\boldsymbol{v}_s(\mathbf{k})  + \boldsymbol{u} _{s}(\mathbf{k}) \,, \quad
\boldsymbol{u}_{s}(\mathbf{k})  =  \nabla_{\mathbf{k}} \zeta _{s}(\mathbf{k})   
= e \, \chi \,v_0 \,  \mathcal{G}_s  \, \frac{ \mathbf{B}  
	-  2\, \hat{ {\mathbf k}} \left (\hat{  {\mathbf k}  } \cdot \mathbf{B} \right) }{k^2}	\,,
\end{align}
where $\xi _s ({\mathbf k})$ and $\boldsymbol{w} _{s}(\mathbf{k})$ are the OMM-modified energy and band velocity of the quasiparticles, respectively. With the usual usage of notations, $\hat{ \mathbf{k} }$ is the unit vector along $ \mathbf{k} $. The full rotational isotropy of the Fermi surface, for each band of the RSW node, is broken by the inclusion of the OMM corrections.

\subsection{Chern numbers}

Using Eq.~\eqref{eqberry}, the Chern number for the $s^{\rm th}$ band can be evaluated as
\begin{align}
\mathcal C_s = \frac{1}{2\,\pi} \int_S d \mathbf S \cdot {\mathbf \Omega}_s \,,
\end{align}
where $S$ denotes a closed surface enclosing the $\mathbf k = \mathbf 0 $ point, and $d \mathbf S$ denotes the outwardly-directed area vector for an infinitesimal patch on $S$. Exploiting the spherical symmetry of the BC flux around an RSW node, we can choose $S$ to be the surface of a unit sphere. Expressing in terms of the spherical polar coordinates,
we use $ k_{x} =  \sin \theta \cos \phi $, $k_{y} =   \sin \theta  \sin \phi  $, and $k_{z}  =   \cos \theta  $, which gives us $ d \mathbf S = \sin \theta \,  d\theta \, d\phi \, {\mathbf {\hat k}} $. This leads to
\begin{align}
\label{eqchern}
\mathcal C_s = - \,\frac{\chi \, s }{2\,\pi} \int_{S} d\theta \,d\phi \sin \theta
=  -\, 2 \,\chi \, s \,.
\end{align}

\section{Quantization observed in nonlinear response}
\label{secnonlinear}

In this section, we will unravel the quantized nature of the ECR and the TCR, associated with the electric and thermal currents, respectively, as explained in the introduction. The reader is referred to Appendix~\ref{secboltz} for a detailed derivation of the forms of the electric and thermal current densities, denoted by ${\mathbf J}_s$ and ${\mathbf J}_s^{\rm th}$, respectively, starting from the Boltzmann equations. The final expressions are obtained by solving for $ \delta  f_s (\mathbf r,\mathbf k)$ [defined in Eq.~\eqref{eqfdsoln}], which is the deviation of the distribution of the quasiparticles from the equilibrium, upto second order in $\epsilon $. Here, $\epsilon \in [0, 1 ] $ is a perturbative parameter, which quantifies the smallness of the magnitude of the probe fields comprising $\mathbf E  $, $\nabla_{\mathbf r} T $, and $\nabla_{\mathbf r} \mu $. In other words, we work in the regime where $ |\mathbf E| \propto \epsilon  $, $ |\nabla_{\mathbf r} T |  \propto \epsilon$, and $ |\nabla_{\mathbf r} \mu | \propto \epsilon$, and use the expression
\begin{align}
f_s (\mathbf r,\mathbf k)
=  f_0  (\varepsilon_s) + \epsilon \, f_{s}^{(1)}  (\varepsilon_s) 
+ \epsilon^2 \, f_{s}^{(2)}  (\varepsilon_s) 
+ \mathcal{O}(\epsilon^3)
\end{align}
in a perturbative expansion. Finally, we set $\epsilon = 1 $ at the end of our computations.

\subsection{Electrochemical response} 
\label{secchem}

For computing the ECR, we set $\mathbf B $ and $\nabla_{\mathbf r} T $ to zero.
The part of the total electrical current density ($ {\mathbf J}_s $), which shows a quadratic dependence on the probe fields $\mathbf E$ and $\nabla_{\mathbf r} \mu $, is denoted by ${\bar {\mathbf J}}^s $. This is the electrochemical current density, with its components being proportional to the quadratic combinations of the form $ E_a  \, \partial_{b} \mu $. Its $a^{\rm th}$ components is given by
\begin{align}
\label{eqelectrocur}
{\bar J}^s_{a}   = \vartheta^s_{a  b c} \, \partial_b\mu \, E_c  \,,
\end{align}
where $\vartheta^s_{a b c }$ is a rank-three conductivity tensor for the $s^{\rm th}$ band (associated with the node of chirality $\chi$). The dependence of the electric current on the the probe fields at second order is the reason why we call the response to be nonlinear.

For computing $ \vartheta^s_{a b c}$, 
we find that the solutions turn out to be
\begin{align}
 f _{s} ^{(1)}  (\varepsilon_s)
=  e \,\tau 
\left( {\boldsymbol v}_s \cdot \boldsymbol{\mathcal {E}} \right )
  f_0^\prime (\varepsilon_s)  \,, \quad
f _{s} ^{(2)}   (\varepsilon_s) = e \,\tau  
\left [ \left ( \boldsymbol{\mathcal {E}} \times \mathbf {\Omega}_{s} 
\right) 
\cdot   {\nabla} _{\mathbf {r}}  \mu \right ]
\, f_0^\prime (\varepsilon_s) \,,
\end{align} 
where
\begin{align}
\boldsymbol{\mathcal E} (\mathbf r) = 
 \mathbf E 
- \frac{ \nabla_{\mathbf r} \mu (\mathbf r) } {e}\, .
\end{align}
Due to the missing of the term $\nabla_{\mathbf r } \mu $ in the Boltzmann equations by the authors in Ref.~\cite{ruiz_electro} [see Eq.~\eqref{eqkin} for more details], our solutions differ from theirs.
Plugging in the solution in Eq.~\eqref{eqcur}, we obtain
\begin{align}
 {\mathbf J}^{s, (2)}
= - \,e^2 \,\tau  \int \frac{d ^{3} \mathbf {k}}{(2 \, \pi) ^{3}}  
\left [ 
\left ( e \,\mathbf E - \nabla_{\mathbf r } \mu \right ) 
\times \mathbf {\Omega} _{s} 
\left( {\boldsymbol v}_s \cdot \mathbf E \right )
 + \boldsymbol{v}_s   
\left \lbrace
\left ( \mathbf E \times \mathbf {\Omega}_{s} 
\right) 
\cdot   {\nabla} _{\mathbf {r}}  \mu \right \rbrace  \right ] 
f_0^\prime (\varepsilon_s)
\end{align}
as the final expression for the response which is second-order in the probe fields, modulo the current density due to the magnetization density [cf. Eq.~\eqref{eqcurtotal}]. From there, we obtain ${\bar {\mathbf J}}^s $, which we divide up as
\begin{align}
{\bar {\mathbf J}}^s  &  
 =  \mathbf{J}^{(s, 1)}   + \mathbf{J}^{(s, 2)} + \mathbf{J}^{(s, 3)}    \,,
\label{eqsecorderf} 
\end{align}
where
\begin{align}
\mathbf{J}^{(s, 1)} &=  e^2 \,\tau 
\int \frac{ d^3 \mathbf k}{(2\, \pi)^3 }  
\left( 
\nabla_{\mathbf r} \mu \cross \mathbf{\Omega}_{s}    \right ) 
\left(  \boldsymbol{v}_{s} \cdot \mathbf{E}   \right )
f^\prime_{0} ( \varepsilon_{s}) \,, \quad
\mathbf{J}^{(s, 2)}  = -\, e^2 \, \tau
\int \frac{ d^3 \mathbf k}{(2\, \pi)^3 }  
\, \boldsymbol{v}_{s} \left[ 
\left( \mathbf{E} \cross \mathbf{\Omega}_{s} \right ) \cdot  
{\nabla}_{\mathbf r} \mu   \right]    f'_{0} (\varepsilon_{s}) \,,\nn
\mathbf{J}^{(s, 3)} &= \frac{ e^2 \, \tau \, \mathcal{G}_{s}}{s^2}
\int  \frac{ d^3 \mathbf k}{(2\, \pi)^3 }  \,   \varepsilon_{s} 
\left( \boldsymbol{v}_s \cdot \mathbf{E} \right) 
\left(  \mathbf{\Omega}_{s} \cross  {\nabla}_{\mathbf r} \mu   \right )   
f^{\prime \prime}_{0} ( \varepsilon_{s} )  \,.
\end{align}
Here, $ \mathbf{J}^{(3)}_{s}  $ represents the part arising from the magnetization density. We include it here, although it does not contribute to transport measurements, to outline its quantized nature (in the absence of tilt).
On analyzing the form of $\mathbf{J}^{(s,3)}$, we find that a nonzero response is obtained when $\mathbf E $ is oriented perpendicular to $\nabla_{\mathbf r} \mu $, and only the component of $\mathbf \Omega_s$, which is perpendicular to both $\mathbf E $ and $\nabla_{\mathbf r} \mu $, will contribute. Now, for an untilted RSW node, we find that $ \boldsymbol{v}_s
\propto \mathbf \Omega_s \propto {\mathbf{\hat k }}$. This immediately tells us that the integral will give a nonzero answer only from the component of $\mathbf \Omega_s$ which is parallel to the currrent density, i.e., $ {J}^{(s, 3)}_a  \propto \epsilon_{a b c } \int d^3 {\mathbf k} \left( v_s \right )_a \left( \Omega_s \right)_a \,\partial_b \mu \, E_c $. An analogous argument for $\mathbf{J}^{(s,1)}$ tells us that $ {J}^{(s,1)}_a  \propto \epsilon_{a b c } \int d^3 {\mathbf k} 
\left( v_s \right )_c \left( \Omega_s \right)_c \,\partial_b \mu \, E_c $.

The corresponding components of the third-rank conductivity tensors are given by
\begin{align}
& \vartheta_{a  b c}^{(s,1)} =  e^{2} \,\tau
\sum_{d}  \epsilon_{a b d }
\int \frac {d^3 \mathbf{k}} {(2 \, \pi)^3 } 
  \left( v_{s} \right)_c
 \left( \Omega_{s} \right)_{d} \, f^{\prime}_{0} (\varepsilon_s)  \,,\quad
\vartheta_{a  b c}^{(s,2)}  = - \,
e^2 \, \tau
\sum_{d}  \epsilon_{b c d}
\int \frac {d^3 \mathbf{k}} {(2 \, \pi)^3 } 
 \left(  v_s \right)_a
\left( \Omega_{s} \right)_d \, f^{\prime}_{0} (\varepsilon_s)  \,,\quad
\nn &  \text{and }
\vartheta_{a  b c}^{(s, 3)}  =  
\frac{ e^2 \, \tau \, \mathcal{G}_{s}} { s^2} 
\sum_{d}  \epsilon_{a d b }
\int \frac {d^3 \mathbf{k}} {(2 \, \pi)^3 } 
\, \varepsilon_{s} \ \left( v_s \right)_c 
 \left( \Omega_s \right)_d  f^{\prime \prime}_{0} (\varepsilon_s)  \,,
\text{ such that } 
\vartheta_{a  b c}^{s} = 
\vartheta_{a  b c}^{(s,1)} + \vartheta_{a  b c}^{(s,2)} +\vartheta_{a  b c}^{(s,3)} \,.
\end{align}
For an untilted RSW node, we find that $\boldsymbol v_s \propto \mathbf \Omega_s \propto {\mathbf{\hat k }}$. This immediately tells us that the integral must be proportional to (1) $\delta_{cd}$ for $\vartheta_{a  b c}^{(s,1)}$  and $\vartheta_{a  b c}^{(s,3)}$; (2) $\delta_{a d}$ for $\vartheta_{a  b c}^{(s,2)}$. This gets rid of the summation over $d$, leading to 
\begin{align}
& \vartheta_{a  b c}^{(s,1)} =  e^{2} \,\tau\,\epsilon_{a b c } 
\, I^{(1)}_{s,c}  \,,\quad
\vartheta_{a  b c}^{(s,2)}  = - \,
e^2 \, \tau\,\epsilon_{a b c } 
 \, I^{(1)}_{s,a}  \,,\quad
 \vartheta_{a  b c}^{(s, 3)}  =   - \,
\frac{ e^2 \, \tau \, \mathcal{G}_{s}} { s^2} \,
\epsilon_{a b c } \, I^{(2)}_{s,c} \,, \quad
\end{align}
where
\begin{align}
I^{(1)}_{s,a}  \equiv  \int \frac {d^3 \mathbf{k}} {(2 \, \pi)^3 } 
\left(  v_s \right)_a
\left( \Omega_{s} \right)_a \, f^{\prime}_{0} ( \varepsilon_s)
\text{ and }
I^{(2)}_{s,a} \equiv  \int \frac {d^3 \mathbf{k}} {(2 \, \pi)^3 } 
\, \varepsilon_s 
\left(  v_s \right)_a
\left( \Omega_{s} \right)_a \, f^{\prime \prime}_{0} (\varepsilon_s) \,.
\end{align}

We employ the coordinate transformations shown in Eq.~\eqref{eqpolar}.
Using Eqs.~\eqref{eqchern}, \eqref{eqinte}, \eqref{equps}, and \eqref{eqinte2}, we conclude that
$ I^{(1)}_{s,x} = I^{(1)}_{s,y} = I^{(1)}_{s,z} = \mathcal I^{(1)}_{s} / 3 $ and
$ I^{(2)}_{s,x} = I^{(2)}_{s,y} = I^{(2)}_{s,z} = \mathcal I^{(2)}_{s} / 3 $
where
\begin{align}
\mathcal I^{(1)}_{s}
=  -\, \frac{ \mathcal C_s} { (2 \, \pi)^2 }  
 \text{ and }
\mathcal I^{(2)}_{s} = -\, \mathcal I^{(1)}_{s}\,.
\end{align}
Plugging these in, we find the final expressions as follows:
\begin{align}
\label{eqfinvartheta}
& \vartheta_{a  b c}^{(s,1)}= - \, \vartheta_{a  b c}^{(s,2)} 
= \frac{\epsilon_{a b c}\,  e^{2} \,\tau}
{3} \,\mathcal I^{(1)}_{s} \,,\quad
 \vartheta_{a  b c}^{(s, 3)}  =  
\frac{ e^2 \, \tau \, \mathcal{G}_{s}} { 3\, s^2} \, 
\epsilon_{a b c }  \, \mathcal I^{(1)}_{s}\,. 
\end{align}
Therefore, each of the response coefficients is proportional to the Chern number, irrespective of the values of $\mu $ and $T$, and, hence, show a completely quantized nature. Also, the nonzero response coefficients are the ones which are proportional to $ \epsilon_{abc} $, representing the condition that $  {\bar {\mathbf J}}^s  \propto 
\mathbf E \cross \nabla_{\mathbf r} \mu $ (i.e., the ones  arising when the components of the electric current vector, $\mathbf E $, and $\nabla_{\mathbf r} \mu $ are oriented orthogonal to each other). It is important to note that for nonzero tilt, although the leading-order terms (on expanding in $ |\eta| \ll 1 $) take the quantized forms as shown above, nonzero subleading terms are generated, which are $\eta $-dependent. A nonzero $\eta$ also generates nonzero components beyond those $\propto  \epsilon_{abc} $.

\subsection{Thermochemical response} 
\label{sectherm}

For computing the TCR, we set $\mathbf B $ and $\mathbf E $ to zero.
The part of the total thermal current density ($ {\mathbf J}_s^{\rm th} $), which shows a quadratic dependence on the probe fields $ \nabla_{\mathbf r} T$ and $\nabla_{\mathbf r} \mu $, is denoted by ${\bar {\mathbf J}}^{{\rm th}, s} $. This is the thermochemical current density, with its components being proportional to the quadratic combinations of the form $ - \,\partial_a T  \, \partial_{b} \mu $. Its $a^{\rm th}$ components is given by
\begin{align}
\label{eqthermocur}
{\bar J}^{\rm th,s}_{a}   = \varphi^s_{a  b c} \, \partial_b\mu 
\left ( -\, \partial_c T  \right) ,
\end{align}
where $\varphi^s_{a b c }$ is a rank-three thermal-conductivity tensor for the $s^{\rm th}$ band (associated with the node of chirality $\chi$). The dependence of the thermal current on the the probe fields at second order is the reason why this response is nonlinear.

For computing $ \var\phi^s_{a b c}$, we find that the solutions turn out to be
\begin{align}
 f_{s} ^{(1)}  (\varepsilon_s)
=  \frac { \tau \left( \varepsilon_s -\mu \right) } {T}
\left( {\boldsymbol v}_s \cdot \nabla_{\mathbf r}  T \right )
  f_0^\prime (\varepsilon_s)  \,, \quad
f _{s} ^{(2)}   (\varepsilon_s) = 
 \frac { \tau \left( \varepsilon_s -\mu \right) } {T}  
\left [ \left ( \nabla_{\mathbf r}  T \times \mathbf {\Omega}_{s} 
\right) 
\cdot   {\nabla} _{\mathbf {r}}  \mu \right ]
\, f_0^\prime (\varepsilon_s) \,.
\end{align} 
Analogous to the ECR case, we get
\begin{align}
{\bar {\mathbf J}}^{{\rm th},s}  &  
 =  \mathbf{J}^{{\rm th},(s, 1)} + \mathbf{J}^{{\rm th},(s, 2)} 
 + \mathbf{J}^{{\rm th},(s, 3)} \,,
\label{eqsecorderf} 
\end{align}
where
\begin{align}
\mathbf{J}^{{\rm th},(s, 1)} & = - \, \frac{\tau}{T} 
\int \frac{d^3 \mathbf{k}}{(2 \, \pi)^3} 
\left ( \varepsilon_s - \mu \right )^2 
\left ( {\nabla}_{\mathbf{r}} \mu \cross \mathbf{\Omega}_s \right )
  \left( \boldsymbol{v}_{s}  \cdot {\nabla}_{\mathbf{r}} T \right) 
 f^{\prime}_0(\varepsilon_s) \,, \nn
\mathbf{J}^{{\rm th},(s, 2)} &= 
 \frac{\tau}{T} 
\int \frac{d^3 \mathbf{k}}{(2 \, \pi)^3} 
\left ( \varepsilon_s - \mu \right )^2 
\left [ 
\boldsymbol{v}_{s} 
 \left ({\nabla}_{\mathbf{r}} T \cross \mathbf{\Omega}_s \right ) 
 \cdot {\nabla}_{\mathbf{r}} \mu 
 \right ] f^{\prime}_0(\varepsilon_s) \,,\nn
\mathbf{J}^{{\rm th},(s, 3)} &= - \, \frac{\tau \, \mathcal{G}_{s}}{s^2 \, T} 
\int \frac{ d^3 \mathbf k}{(2\, \pi)^3 }  
\left ( \varepsilon_s - \mu \right ) \varepsilon_s  
\left( \boldsymbol{v}_{s} 
\cdot   \, {\nabla}_{\mathbf r} T   \right )  
 \left(  \mathbf{\Omega}_s \cross {\nabla}_{\mathbf{r}} \mu \right ) 
\left[ ( \varepsilon_s - \mu ) \,  f^{\prime \prime}_{0} (\varepsilon_s) 
 + f^{\prime}_{0} (\varepsilon_s ) \right]  .
\end{align}
Here, $ \mathbf{J}^{{\rm th},(s, 3)}  $ represents the part arising from the magnetization density.

The corresponding components of the third-rank thermal conductivity tensors are given by
\begin{align}
\varphi_{a  b c}^{(s,1)}  & = \frac{ \tau  }
{  T }   \sum_{d}
 \int  \frac{ d^3 \mathbf{k} }
{ (2 \,\pi)^{3} }
 \, \left  ( \varepsilon_s - \mu  \right )^2  \epsilon_{ a b d } 
 \left( v_s \right)_c \left(  \Omega_s \right)_d
 f^{\prime}_{0} (\varepsilon_s ) \,, \nn
 \varphi_{a  b c}^{(s,2)} & =  -\, \frac{ \tau  }
{  T }   \sum_{d}
\int 
\frac{ d^3 \mathbf{k} }
{(2 \, \pi)^3} \left ( \varepsilon_s - \mu \right )^2
 \, \epsilon_{ b c d} 
\left(  v_{s} \right)_a  
\left( \Omega_s \right)_d f^{\prime}_{0} (\varepsilon_s ) \,, \text{ and} \nn
  \varphi_{a  b c}^{(s, 3)}  & =  
\frac{ \tau \,  \mathcal{G}_s } { \, T \, s^2}  \sum_{d}
\int \frac{ d^3 \mathbf{k}}
{(2 \pi)^{3} }
 \left( \varepsilon_s   - \mu \right )  
\varepsilon_s \,
\epsilon_{a d b } \left( v_s \right)_c \left( \Omega_s \right)_d 
 \left[ ( \varepsilon_s - \mu ) \,  
 f^{\prime \prime}_{0} (\varepsilon_s) + f^{\prime}_{0} (\varepsilon_s ) \right],
\nn 
\text{such that } 
\varphi_{a  b c}^{s}  & =
\varphi_{a  b c}^{(s,1)} + \varphi_{a  b c}^{(s,2)} +\varphi_{a  b c}^{(s,3)} \,.
\end{align}
The relations $\boldsymbol v_s \propto \mathbf \Omega_s \propto {\mathbf{\hat k }}$ for an untilted node again constrain the integrals to be proportional to (1) $\delta_{cd}$ for $\varphi_{a  b c}^{(s,1)}$  and $\varphi_{a  b c}^{(s,3)}$; (2) $\delta_{a d}$ for $\varphi_{a  b c}^{(s,2)}$. Hence, we get rid of the summation over $d$, leading to
\begin{align}
\varphi_{a  b c}^{(s,1)}   = \frac{ \tau \, \epsilon_{ a b c }   }
{  T }  \, I^{(3)}_{s,c} \,, \quad
 \varphi_{a  b c}^{(s,2)} & =  -\, \frac{ \tau \, \epsilon_{ a b c} }
{  T }  \, I^{(3)}_{s,a} \,, \quad
  \varphi_{a  b c}^{(s, 3)}   =  -\,
\frac{ \tau \,  \mathcal{G}_s \, \epsilon_{a b c }} 
{ \, T \, s^2} \, I^{(4)}_{s,c} \,.
\end{align}
where
\begin{align}
I^{(3)}_{s,a}  \equiv  \int \frac {d^3 \mathbf{k}} {(2 \, \pi)^3 } 
\left  ( \varepsilon_s - \mu  \right )^2 
\left(  v_s \right)_a
\left( \Omega_{s} \right)_a \, f^{\prime}_{0} (  \varepsilon_s)
\text{ and } 
I^{(4)}_{s,a}  \equiv  \int \frac{ d^3 \mathbf{k}}
{(2 \pi)^{3} }  \,
\varepsilon_s \left( v_s \right)_a \left( \Omega_s \right)_a 
 \left[ \left ( \varepsilon_s - \mu \right )^2 
 f^{\prime \prime}_{0} (\varepsilon_s) 
+ \left ( \varepsilon_s - \mu \right )  f^{\prime}_{0} (\varepsilon_s ) \right].
\end{align}
Employing  Eqs.~\eqref{eqpolar}, \eqref{eqchern}, \eqref{eqinte}, \eqref{equps}, and \eqref{eqinte2}, we conclude that
$ I^{(3)}_{s,x} = I^{(3)}_{s,y} = I^{(3)}_{s,z} = \mathcal I^{(3)}_{s} / 3 $ and
$ I^{(4)}_{s,x} = I^{(4)}_{s,y} = I^{(4)}_{s,z} = \mathcal I^{(4)}_{s} / 3 $
where
\begin{align}
\mathcal I^{(3)}_{s}  = -\, \frac{ \pi^2 \, T^2 \,\mathcal C_s} 
{ 3 \times (2 \, \pi)^2 } 
 = \frac{\pi^2 \, T^2} {3} \, \mathcal I^{(1)}_{s}
\text{ and }
\mathcal I^{(4)}_{s}
= -\, \frac{ 2\, \pi^2 \, T^2} {3} \,\mathcal I^{(1)}_{s}\,.
\end{align}

Plugging these in, we find the final expressions as follows:
\begin{align}
\label{eqfinvarphi}
& \varphi_{a  b c}^{(s,1)}= -\varphi_{a  b c}^{(s,2)} 
= \epsilon_{a b c}\, \frac{\pi^2 \, T  } {3}  \, \tau\, \mathcal I^{(1)}_{s}
 \,,\quad
 \varphi_{a  b c}^{(s, 3)}  =   \frac{ 2\, \pi^2 \, T  } {3} \,
 \frac{  \tau \, \mathcal{G}_{s}} { s^2} \, 
\epsilon_{a b c }  \, \mathcal I^{(1)}_{s} \,. 
\end{align}
Clearly, the results are quantized only at a fixed value of $T$, as the overall response increases linearly with increasing $T$. Needless to say here that the nonzero response coefficients are the ones which are proportional to $ \epsilon_{abc} $, representing the condition that $  {\bar {\mathbf J}}^{{\rm th}, s}  \propto 
\nabla_{\mathbf r} T \cross \nabla_{\mathbf r} \mu $ (i.e., the ones  arising when the components of the thermal current vector, $\nabla_{\mathbf r} T $, and $\nabla_{\mathbf r} \mu $ are oriented orthogonal to each other).
Similar to the ECR case, for a nonzero tilt, in addition to the leading-order terms (on expanding in $|\eta| \ll 1 $) taking the quantized forms as shown above, $\eta $-dependent terms are generated. A nonzero $\eta$ will also be responsible for giving rise to nonzero components beyond those $\propto  \epsilon_{abc} $.

From Eqs.~\eqref{eqfinvartheta} and \eqref{eqfinvarphi}, we find that
\begin{align}
\vartheta_{a  b c}^{(s,i)}=
 \frac{ 3 \,e^2 } {\pi^2\, T}   \, \varphi_{a  b c}^{(s,i)}  
\label{eqwf}
\end{align}
is satisfied for $i=1$ and $i=3$, which embodies the Wiedemann-Franz law \cite{mermin}.
However, $\vartheta_{a  b c}^{(s, 3)}=
 \frac{ 3 \,e^2 } { 2\,\pi^2\, T}   \, \varphi_{a  b c}^{(s, 3)}  $, which are the parts arising from the magnetization density. This is fine because the Wiedemann-Franz law is primarily applicable for linear response, and may not be universally valid in the nonlinear regimes (see, for example~\cite{nandy-wf}) or for the magnetization currents. A few more comments are in order: Let us define the scattering probability ($ W_{\mathbf k, \mathbf q}$) as the probability that a quasiparticle, with momentum $\mathbf k $, is scattered into any of the quantum states contained in the infinitesimal volume $d^3 \mathbf q$, centred around $\mathbf q $ (assuming those states are unoccupied in order to satisfy the Pauli exclusion principle).
At sufficiently low temperatures, scatterings caused by impurities are the dominant source in any realistic material \cite{mermin}. If the impurity-concentration is sufficiently low and the electron-impurity coupling is sufficiently weak, the actual derivation of the collision integral shows that the scattering is necessarily inelastic. This leads to the scattering probabality $W_{\mathbf k, \mathbf q}$ to be essentially zero unless $\varepsilon_s (\mathbf k) =  \varepsilon_s (\mathbf q)$. Under these circumstances, the Wiedemann-Franz law holds \cite{mermin}. However, for inelastic scattering processes, when we can have a nonvanishing $ W_{\mathbf k, \mathbf q}$ for $\varepsilon_s (\mathbf k) \neq  \varepsilon_s(\mathbf q) $, the Wiedemann-Franz law is not expected to hold generically. However, such detailed analysis can be performed only by going beyond the relaxation-time approximation, which is beyond the scope of work --- hence, we leave it for future investigations.

\section{Magnetoelectric conductivity} 
\label{sechall}

In a planar Hall set-up, comprising static uniform electric and magnetic fields applied parallel to each other, a nonzero tilt induces the linear response from an RSW node to contain parts which vary linearly with $B$ (which we demonstrate below). In this section, our aim is to elucidate how the characteristic topological quantities arise in those linear-in-$B$ parts of the magnetoconductivity tensor.

In the planar Hall configuration, the magnetoelectric conductivity tensor is obtained from Eq.~\eqref{eqcur}, with the solution
(see Ref.~\cite{ips-rsw-ph} for detailed derivations starting from the Boltzmann equations)
\begin{align} 
\label{eqbolinter}
f_{s} (\mathbf{k})   &=  f_{0} (\xi_{s}) +\delta f_{s} (\mathbf{k}) \,,
\quad
\delta f_{s} (\mathbf{k}) 
 = \left [- f_{0}^\prime (\xi_{s}) \right ]  {\tilde g}_s  (\mathbf{k})\, ,
 \quad
\frac{ {\tilde g}_s (\mathbf{k}) } 
{ e \, \tau }
= -\, \sum_{n = 0}^{\infty}
\left (e \, \tau \, \mathcal{D}_{s} \right )^n \hat{L}^n 
\left [    \mathcal{D}_{s} \,
 \left ( \boldsymbol{w}_{s} 
+ \boldsymbol{W}_s \right ) \cdot \mathbf{E}   \right ], 
\end{align} 
where
\begin{align}
\nn &
\mathbf{W}_s =  
 e \left ( \boldsymbol{w}_s  \cdot \boldsymbol{\Omega}_s
 \right ) \mathbf{B} \,, \text{ and }
\hat{L} = (\boldsymbol{w}_s \cross \mathbf{B}) \cdot \nabla_{\mathbf{k}}
\end{align} 
is referred to as the Lorentz-force operator (because it includes the classical effect due to the Lorentz force).
We would like to point out here that, in Ref.~\cite{ips-rsw-ph}, two of us have studied the question of linear response for planar Hall and planar thermal Hall set-ups consisting of untilted RSW nodes (i.e., for $\eta = 0$). As such, the response coefficients did not have any part with a linear-in-$B$ dependence. The fact that the linear-in-$B$ parts go to zero for $\eta = 0$ \cite{ips-rsw-ph} is expected because, according to the Onsager-Casimir reciprocity
relations \cite{onsager31_reciprocal, onsager2, onsager3, rahul-jpcm}, terms containing odd powers of $B$ cannot arise in this situation. However, in the presence of a nonzero tilt, the conditions are relaxed, making it possible for the existence of terms varying linearly with $B$ \cite{rahul-jpcm}.

Here, we will consider the case when $\boldsymbol{E}$ and $\boldsymbol{B}$ are parallel to each other (see Ref.~\cite{li-quantize} for the same set-up applied to WSMs). Therefore, if $ \mathbf E $ is applied along the unit vector
$ \hat{\mathbf e}_E $, such that $\mathbf E = E  \,\hat{\mathbf e}_E$, we have $\mathbf B =  B \, \hat{\mathbf e}_E $. Note that, if $\mathbf B = B \,{\mathbf{\hat b}} $, $ \zeta _{s} 
 = - \left(  m _s  \right )_b \,  B
=  e \, \chi\, v_0 \, \mathcal{G}_s  \,   
\frac{  k_b\, B } {k^2}  $ [cf. Eq.~\eqref{eqomme}]. We are primarily interested in finding the form of the part of the magnetoelectric conductivity tensor, denoted by $\left( \sigma_s^{\rm lin} \right)_{a  b} $, which is linear-in-$B$.
Expressing the corresponding part of the current as
$
J^{{\rm lin}, s}_{a} = \left( \sigma_s^{\rm lin} \right)_{a  b} \, E^b  \,,$
let us define a third-rank tensor
\begin{align}
\left( \varsigma_s^{\rm lin} \right )^c_{\; \;a b} 
= \frac{\partial \left( \sigma_s^{\rm lin} \right)_{a  b} }
{\partial B_c} \,.
\end{align}
For our scenario with $\mathbf E \parallel \mathbf B $, we will be dealing with the vector defined as
\begin{align} 
\varsigma^s_{a}=
\sum_{b, c} \left( \varsigma_s^{\rm lin} \right )^c_{\; \;a b}  \,\delta^b_{\;\; c} 
= \sum_{b} \left( \varsigma_s^{\rm lin} \right )^b_{\; \;a b}
\equiv \varsigma_{a}^{(s,1)}+\varsigma_{a}^{( s, 2)}
+ \varsigma_{a}^{(s, 3)} + \varsigma_{a}^{(s, 4)} +\varsigma_{a}^{(s, 5)}  \,,
\end{align}
where (see Appendix~\ref{appsigma} for the intermediate steps)
\begin{align}
& \varsigma_{a}^{( s, 1)} 
= e^3 \, \tau 
 \int \frac{ d^3 \mathbf{k}}
 {(2 \, \pi)^3}
  \left( v_s \right)_a \left (\boldsymbol{v}_{s} \cdot \mathbf{\Omega}_{s} \right )   
  f^{\prime}_{0} (\varepsilon_s) 
 =  -\, \frac{ \delta_{z a} \, \eta\, v_0 \,  e^3 \, \tau 
\, \mathcal{C}_s }
{ 4 \, \pi^2   }   \,,\quad
\varsigma_{a}^{(s, 2)}   = - \, 4 \, \varsigma_{a}^{( s, 1)} \,,\nn &
\varsigma_{a}^{(s, 3)}   =  \frac{e^3\, \tau \, v_0 \, \mathcal{G}_s} {s}
\int \frac{d^3 \mathbf{k} }
{(2 \, \pi)^3} \left( v_s \right)_a 
\left [\nabla_{\mathbf k}  \cdot 
\left( k\, \mathbf \Omega_s \right) \right ]
f^{\prime}_{0} (\varepsilon_s)
=  -\, \frac{
\delta_{z a} \, \chi \, v_0 \, e^3\, \tau  \, \mathcal{G}_s
} { 2\, \pi^2 } \,
\frac {\left (s^2 - \eta^2 \right)
\tanh^{-1}\big (\frac {\eta} {s} \big) - \eta  \, s } 
{\eta^2} \text{ for } |\eta| < |s| \,, 
\nn \varsigma_{a}^{(s, 4)}  &
 =  \frac{e^3\, \tau \, v_0 \, \mathcal{G}_s} {s} 
\int \frac{d^3 \mathbf{k} }
{(2 \, \pi)^3} \left [  \boldsymbol v_s \cdot  
  \partial_{k_a } (k \, \mathbf \Omega_s ) \right ]
 f^{\prime}_{0} (\varepsilon_s) = \varsigma_{a}^{(s, 3)} \,,
\quad
\varsigma_{a}^{(s, 5)}   
= \frac{e^3\, \tau \, v_0 \, \mathcal{G}_s} {s}
 \int \frac{d^3 \mathbf{k} } 
 {(2 \, \pi)^3}
\left( v_s \right)_a k
\left( \boldsymbol v_s \cdot 
\mathbf  \Omega_s \right) f^{\prime \prime}_{0} (\varepsilon_s)
= -\, \varsigma_{a}^{(s, 3)}\,.
\end{align}  
Appendix~\ref{secfinvarsig} provides more details regarding the steps leading to the final expressions shown above.
Here, $\varsigma_{a}^{( s, 1)}$ and $\varsigma_{a}^{( s, 2)}$ represent the parts arising purely from the BC (i.e., with no contribution from the OMM), $\varsigma_{a}^{( s, 3)}$, $\varsigma_{a}^{( s, 4)}$, and $\varsigma_{a}^{( s, 5)}$ are the terms which go to zero if OMM is neglected.
\begin{figure}[t]
\centering
\includegraphics[width=0.45 \textwidth]{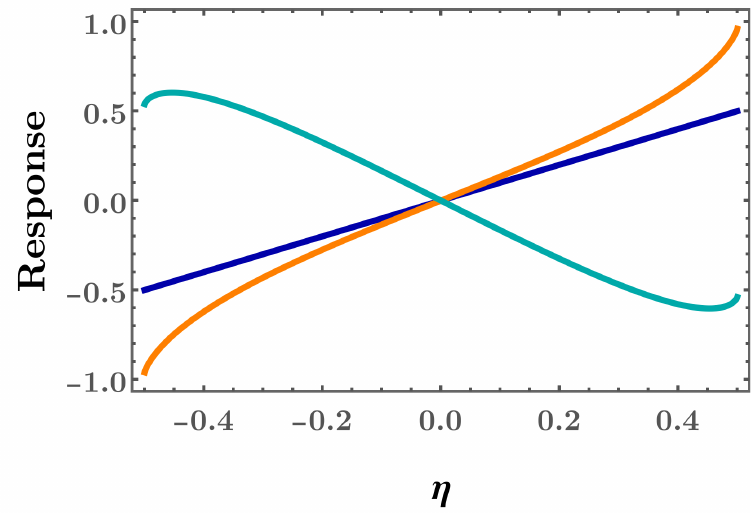}
\caption{\label{figsigma}
Plots for $ \mathcal F_1 \equiv
\sum \limits_{s=1/2, 3/2}
\frac {- \,\eta \, \mathcal{C}_s } {4} $ (blue line), $
\mathcal F_2 \equiv
\sum \limits_{s=1/2, 3/2} 
\frac{ \chi \, \mathcal{G}_s
} { 2 } \,
\frac { \eta  \, s -
\left (s^2 - \eta^2 \right)
\tanh^{-1}\big (\frac {\eta} {s} \big)  } 
{\eta^2}$ (orange curve), and $ \mathcal F_{\rm tot} \equiv \mathcal F_1  - 4\,  \mathcal F_1 +  \mathcal F_2 $ (cyan curve) as functions of $\eta $, after setting $\chi= +1$. We note that $\mathcal F_{\rm tot}$ corresponds to the the sum
$\sum \limits_{a=1}^4 \sum \limits_{s=1/2, 3/2} \varsigma_a^s $. The behaviour for $\chi= -1$ is obtained simply as the negative of the above curves.
}
\end{figure}
Some comments are in order:
\begin{enumerate}

\item
We find that all the above response coefficients go to zero for $\eta = 0 $, validating our arguments that the linear-in-$B$ terms do not survive in the limit of zero tilt. 

\item Since $\varsigma_{a}^{( s, 1)} $ and $ \varsigma_{a}^{( s, 2)}$ are proportional to $\eta\, \mathcal C_s$, not depending on $\mu$ and $ T $, they reflect a robust quantized response. Hence, it is easy to identify this part of the response irrespective of the specific material one chooses to study in experiments. The corresponding linear-in-$\eta$ behaviour is illustrated in Fig.~\ref{figsigma} (blue line).
 
\item Although $\varsigma_{a}^{( s, 3)} $, $ \varsigma_{a}^{( s, 4)}$, and $ \varsigma_{a}^{( s, 5)}$ are proportional to $ \chi \, \mathcal{G}_s $, and are independent of $\mu$ and $ T $, they have a complicated (nonlinear) dependence on $\eta$. Since $\eta $ is a nonuniversal parameter, changing its value according to the material under consideration, it is then hard to disentangle/extract the feature of $\mathcal{G}_s$-proportionality. The corresponding nonlinear behaviour is illustrated in Fig.~\ref{figsigma} (orange curve).

\end{enumerate}
Despite the fact that it is difficult to disentangle and decipher the presence of $\mathcal C_s $ and $\mathcal{G}_s$ in the magnetoconductivity in real experiments, our motivation here is strictly to understand theoretically how these quantities determine the net response. In other words, we believe that our theoretical results illustrate how topological semimetals have the signatures of topology robustly embedded in the response functions.

\section{Summary and future perspectives} 
\label{sec_summary}

In this paper, we have considered the response coefficients in two distinct experimental set-ups, with the intent of identifying clear signatures of the topological features of the bandstructures of nodal-point semimetals, by taking RSW semimetal as an example. In the first case, the response is nonlinear in nature, arising under the combined effects of an external electric field (or temperature gradient) and the gradient of the chemical potential. When internode scatterings can be ignored, the relaxation-time approximation leads to the ECR and TCR taking quantized values for a single untilted RSW node, insensitive to the specific values of the chemical potential and the temperature. The second case investigates the linear-response coefficient in the form of the magnetoelectric conductivity
under the action of collinear electric and magnetic fields, resulting in a planar Hall set-up. In particular, we have considered the nature of the linear-in-$B$ parts of the conductivity tensor, which are nonvanishing only in the presence of a nonzero tilt. We find that the components show a quantized nature with coefficients which are functions of $ \eta $.
In our calculations, we have assumed the same relaxation times to be applicable for all the bands.

The advantage of considering RSW semimetals is that it provides a richer structure for demonstrating the topological features contained in the response coefficients, compared to the WSMs, because of the fact that the former consists of four bands (rather than just two). Basically, each band has its distinct topological features, reflected by the band-dependent values of the BC and the OMM. Through our band-dependent results for the relevant response coefficients, we have, thus, succeeded in chalking out the signatures of these individual bands. In the future, it will be worthwhile to calculate the behaviour of the nonlinear response coefficients in situations where internode scatterings are included too \cite{ips-internode}. 
Furthermore, we would like to improve our calculations by going beyond the relaxation-time approximation, which involves actually computing the collision integrals for all relevant scattering processes \cite{timm} (involving both intraband and interband interactions), rather than just using phenomenological values of momentum-independent relaxation times. This will help us make quantitative predictions for realistic scenarios, because disorder-induced scatterings will give rise to all possible generic processes. Last but not the least, it will be interesting to investigate the effects of anisotropy/disorder \cite{ips-cd1, ips_cd, ips-kush} and/or strong interactions on the quantized nature of the response \cite{ips_cpge}.

 \appendix
 
\section{Response resulting from the Boltzmann equations}
\label{secboltz}

In this appendix, we will outline the semiclassical Boltzmann equation (BE) formalism \cite{ips-kush-review, ips_rahul_ph_strain, ips-rsw-ph}, which is used here for determining the transport coefficients. This framework is applicable in the regime of small deviations from the equilibrium quasiparticle distribution. If there is an externally-applied magnetic field $\mathbf B$, then we will we assume that it is small in magnitude, associated with a small cyclotron frequency $\omega_c=e\,B/(m^*\, c) $ (where $m^* $ is the effective mass with the magnitude $\sim 0.11 \, m_e$ \cite{params2}, with $m_e$ denoting the electron mass). This condition is necessary when we want to focus on the regime where quantized Landau levels need not be considered, given by the inequality $\hbar \, \omega_c \ll \mu$, where $\mu$ is the chemical potential cutting the energy band(s), thus defining the Fermi-energy level. Here, we will approximate the form of the collision integral with a relaxation time ($ \tau $), which is a momentum-independent phenomenological parameter. Furthermore, we will focus on the limit when only intranode and intraband scatterings dominate in the collision processes, such that $\tau$ corresponds exclusively to the intranode scattering time.\footnote{We would like to mention here that the effects of internode scatterings in the linear response for multifold semimetals have been addressed in Ref.~\cite{ips-internode}, using again the simple relaxation-time approximation.} Under these approximations, it is sufficient to derive the relevant expressions for a single node, whose chirality is denoted by $\chi$. We note that the neglect of interband scatterings is justified if only those scattering processes are allowed which preserve pseudospin.

For a 3d system, we define the distribution function $ f_s  ( \mathbf r , \mathbf k, t) $ for the quasiparticles occupying a Bloch band labelled by the index $s$, with the crystal momentum $\mathbf k$ and dispersion $\varepsilon_s (\mathbf k)$, such that
$
dN_s = g_s \,f_s ( \mathbf r , \mathbf k, t) \,
\frac{ d^3 \mathbf k}{(2\, \pi)^3 } \,d^3 \mathbf r
$
is the number of particles occupying an infinitesimal phase space volume of $ dV_p = \frac{ d^3 \mathbf k}{(2\, \pi)^3 } 
\,d^3 \mathbf r $, centered at $\left \lbrace \mathbf r , \mathbf k \right \rbrace $ at time $t$. Here, $g_s$ denotes the degeneracy of the band.
We define the OMM-corrected dispersion and the corresponding modified Bloch velocity as
\begin{align} 
\label{eqtoten1}
	\xi _s ({\mathbf k})  = \varepsilon_{s}(\mathbf{k})  +  \zeta _{s}(\mathbf{k})   
\text{ and } 
\boldsymbol{w} _{s}(\mathbf{k})= 
 \nabla_{\mathbf{k}}   \varepsilon_{s}(\mathbf{k})  + \nabla_{\mathbf{k}}\zeta _{s}(\mathbf{k})\,,
\end{align}
respectively. 

 It is convenient to introduce a combined electrochemical potential $ \Phi(\mathbf r) 
+ \frac{\mu (\mathbf r) } {e} $, giving rise to a generalized (external) force field defined by
\begin{align}
\boldsymbol{\mathcal E} (\mathbf r) = 
-\nabla_{\mathbf r} \left [  \Phi(\mathbf r) + \frac{\mu (\mathbf r) } {e}
\right ] ,\,\quad \mathbf E = -\nabla_{\mathbf r } \Phi \,,
\end{align}
where $\Phi (\mathbf r) $ is the scalar potential.
The Hamilton's equations of motion for the quasiparticles, under the influence of a static electrochemical potential and a static magnetic ($\mathbf{B}$) field, are given by \cite{mermin, sundaram99_wavepacket, li2023_planar}
\begin{align}
\label{eqrkdot}
\dot {\mathbf r} &= \nabla_{\mathbf k} \, \xi _{s}  
- \dot{\mathbf k} \, \cross \, \mathbf \Omega _{s} 
 \text{ and }  
\dot{\mathbf k} = -\, e  \left( \boldsymbol {\mathcal E}  +  
	\dot{\mathbf r} \, \cross\, {\mathbf B} 
	\right ) \nn
\Rightarrow  \, 	\dot{\mathbf r}  & = \mathcal{D} _{s} 
	\left[   \boldsymbol{w} _{s} + e 
\left	(  \boldsymbol {\mathcal E}  \cross   \mathbf \Omega _{s} \right )  
+ e   \, ( \mathbf \Omega _{s} \cdot 
\boldsymbol{w} _{s} ) \, \mathbf B  \right] \text{ and }
\dot{\mathbf k}  = -\, e \,\mathcal{D} _{s} 
\left[    \boldsymbol {\mathcal E} 
+  \left (      \boldsymbol{w} _{s}   \cross  {\mathbf B} \right ) 
+  e  \left (    \boldsymbol {\mathcal E}  \cdot  {\mathbf B} \right )  
\mathbf \Omega _{s}  \right].
\end{align}
where $-\, e$ is the charge carried by each quasiparticle.
Furthermore,
\begin{align}
\mathcal{D} _{s} =  \frac{1}
{ 1 + \, e  \left( \mathbf{B} \cdot\mathbf \Omega _{s}\right) }
\end{align} 
is the factor which modifies the phase volume element from $dV_p $ to $  (\mathcal{D} _{s})^{-1} \, dV_p$, such that the Liouville’s theorem (in the absence of collisions) continues to hold in the presence of a nonzero BC \cite{son13_chiral, xiao05_berry, duval06_Berry, son12_berry}.
Putting everything together, the BE for the quasiparticles takes the form \cite{lundgren14_thermoelectric, amit_magneto}:
\begin{align}
\label{eqkin}
& \mathcal{D} _{s} \,\left [ \partial_t  
+ \left \lbrace  \boldsymbol{w} _{s}
+ e 
\, \boldsymbol {\mathcal E} \cross \mathbf{\Omega} _s
+ e \left(  {\mathbf \Omega} _{s} \cdot \boldsymbol{w} _{s}  \right)
	\mathbf{B} \right \rbrace 
	\cdot \nabla_{\mathbf r} 
-  e \left(  \boldsymbol {\mathcal E} + \boldsymbol{w} _{s}  \cross {\mathbf B} 	\right) 
	\cdot \nabla_{\mathbf k} 
- e^2 \left ( \boldsymbol {\mathcal E} \cdot {\mathbf B} \right )
{\boldsymbol{\Omega}}_s   \cdot \nabla_{\mathbf k} \right ] f_s  = I_{\rm coll}\,.
\end{align}
which results from the Liouville’s equation in the presence of scattering events. On the right-hand side, $I_{\rm coll}$
denotes the collision integral, which corrects the Liouville’s equation, taking into account the collisions of the quasiparticles.
We would like to point out that the term represented by $\nabla_{\mathbf r} \mu$ was missed in Ref.~\cite{ruiz_electro}, where the authors attempted to demonstrate the quantized ECR in Weyl semimetals.

Let the contributions to the average DC (generalized) electric and thermal current densities from the quasiparticles, associated with the band $s$ at the node with chirality $\chi$, be ${\mathbf J}_s $ and ${\mathbf J}^{{\rm th}}_s$, respectively. The response matrix, which relates the resulting generalized current densities to the driving electrochemical potential gradient and temperature gradient, is expressed as
\begin{align}
\label{eqcur1}
\begin{bmatrix}
\left( J_s \right)_a \vspace{0.2 cm} \\
\left( {J}_s^{{\rm th}}\right)_a 
\end{bmatrix} & = \sum \limits_b
\begin{bmatrix}
 \left( \sigma_{s}  \right)_{ a b} &  \left( \alpha_{s}  \right)_{ a b}
\vspace{0.2 cm}  \\
T  \left( \alpha_{s} \right)_{ a b } &  \left( \ell_{s}  \right)_{ a b}
\end{bmatrix}
\begin{bmatrix}
\mathcal E_b
\vspace{0.2 cm}  \\
- \, { \partial_{b} T } 
\end{bmatrix} ,
\end{align}
where $ \lbrace a, b \rbrace  \in \lbrace x,\, y, \, z \rbrace $ indicates the Cartesian components of the current density vectors and the response tensors in 3d.
The symbols $\sigma_s $ and $\alpha_s$ represents the magnetoelectric conductivity and the
magnetothermoelectric conductivity tensors, respectively. The latter determines the Peltier ($\Pi_s$), Seebeck ($ S_s $), and Nernst coefficients. The third tensor $\ell_s $ represents the response relating the thermal current density to the temperature gradient, at a vanishing electric field. $ S_s $ , $\Pi_s $, and the magnetothermal coefficient tensor $\kappa_s $ (which provides the coefficients between the heat current density and the temperature gradient at vanishing electric current) are related as \cite{mermin, ips-kush-review}:
\begin{align}
\label{eq:kappa}
\left( S_{s}  \right)_ {ab} = \sum \limits_{ i^\prime}
\left( \sigma_s \right)^{-1}_{ a a^\prime }
\left ( \alpha_s\right)_{ a^\prime b} \, , \quad
 \left( \Pi_{s} \right)_ {ab} = T \sum \limits_{ a^\prime}
\left( \alpha _s \right)_{ a   a^\prime}   
\left(\sigma _s \right)^{-1}_{ a^\prime b} \,,\quad 
\left( \kappa_{s}  \right)_ {a b} =
\left( \ell _s \right)_{ a b }
- T \sum \limits_{ a^\prime, \, b^\prime }
\left( \alpha _s \right)_{ a  a^\prime }
\left( \sigma_s  \right)^{-1}_{  a^\prime  b^\prime }
\left( \alpha _s \right)_{ b^\prime b}  \,.
\end{align}
Since $\ell_s  $ determines the first term in the magnetothermal coefficient tensor $\kappa_s  $, here we will loosely refer to $ \ell_s $ itself as the magnetothermal coefficient.

The explicit form of the DC charge-current current density is given by \cite{xiao06_berry, xiao_review, das-agarwal_omm}
\begin{align}
\label{eqcurtotal}
{\mathbf J}_s^{\rm tot} 
& =  - \,e   \,g_s \int
\frac{ d^3 \mathbf k}{(2\, \pi)^3 } \,
\mathcal{D}_s^{-1}  \,\dot{\mathbf r} \, f_s ( \mathbf r , \mathbf k) 
+   \mathbf {\nabla} _{\mathbf{r}} \times \mathbf{M}_s (\mathbf r)\,,\quad
\mathbf{M}_s (\mathbf r) = g_s \int
\frac{ d^3 \mathbf k}{(2\, \pi)^3 } \,
\mathcal{D}_s^{-1}    \, \mathbf {m} _{s} (\mathbf {k})
\, f_s ( \mathbf r , \mathbf k) \,,
\end{align}
where $ \mathbf{M}_s (\mathbf r) $ is the magnetization density.
As discussed earlier, $ \mathbf {m}_{s} (\mathbf {k}) $ represents the the orbital magnetic moment (OMM), which generically describes the rotation of a wavepacket around its center of mass. The contribution from the magnetic moments to the local current density must be subtracted out in the transport current, because the magnetization current cannot be measured by conventional transport experiments \cite{cooper-halperin}. Therefore, the transport current is defined by $ {\mathbf J}_s =  {\mathbf J}_s^{\rm tot} 
-  g_s  \mathbf {\nabla} _{\mathbf{r}} \times
 \int
\frac{ d^3 \mathbf k}{(2\, \pi)^3 } \,
\mathcal{D}_s^{-1}    \, \mathbf {m}_{s} (\mathbf {k})
\, f_s ( \mathbf r , \mathbf k) $, leading to
\begin{align}
{\mathbf J}_s  =  - \,e   \,g_s \int
\frac{ d^3 \mathbf k}{(2\, \pi)^3 } \,
\mathcal{D}_s^{-1}  \,\dot{\mathbf r} \, f_s ( \mathbf r , \mathbf k) \,.
\end{align}
In an analogous way, the transport heat current is captured by \cite{xiao06_berry}
\begin{align}
\mathbf{J}^{\rm th}_s = g_s  \int
\frac{ d^3 \mathbf k}{(2\, \pi)^3 } 
\, (\mathcal{D}_{s})^{-1}   
\, \dot{\mathbf r} \,
\left( \xi_s - \mu \right)  f_s ( \mathbf r , \mathbf k) \,.
\end{align}
Plugging in the expressions shown in in Eq.~\eqref{eqrkdot}, we get
\begin{align}
\label{eqcur}
{\mathbf J}_s
& =   -\, e \,  g_s  \int
\frac{ d^3 \mathbf k}{(2\, \pi)^3 } \,
\left[   \boldsymbol{w}_{s} 
+ e \left ( \boldsymbol {\mathcal E}  \cross  \mathbf \Omega_{s} \right)  
+ e   \left  ( \mathbf \Omega_{s} \cdot  \boldsymbol{w}_{s} \right )  \mathbf B  \right]
\,  f_s( \mathbf r , \mathbf k) \nn  \text{and } 
 \mathbf{J}^{\rm th}_s  & = g_s  \int
\frac{ d^3 \mathbf k} {(2\, \pi)^3 } 
\left[   \boldsymbol{w}_{s} 
+ e \left ( \boldsymbol {\mathcal E}  \cross  \mathbf \Omega_{s} \right) 
+ e   \left ( \mathbf \Omega_{s} \cdot 
	  \boldsymbol{w}_{s} \right )  \mathbf B  \right]
\left( \xi_s - \mu \right)  f_s ( \mathbf r , \mathbf k)\,.
\end{align}

Under the relaxation-time approximation, the collision integral takes the form of
\begin{align}
I_{\rm coll} =   
	\frac{ f^{(0)}_{s} (\mathbf r,\mathbf k)- f_s (\mathbf r,\mathbf k, t) }
	{\tau } \,,
\end{align}
where the time-independent distribution function
\begin{align}
	f^{(0)}_{s} (\mathbf r,\mathbf k) \equiv 
	f_0 \big (\xi _s(\mathbf k) , \mu(\mathbf{r}), T (\mathbf r) \big )
= \frac{1}
{ 1 + \exp [ \frac{ \xi _s(\mathbf k)-\mu (\mathbf{r})} 
	{ T  (\mathbf r )}  ]}\,,
\end{align} 
describes a local equilibrium situation at the subsystem centred at position $\mathbf r$, at the local temperature $T(\mathbf r )$, and with a local chemical potential $\mu (\mathbf{r}) $. The gradients of the equilibrium distribution function evaluate to
\begin{align}
{\nabla}_{\mathbf k} \, f_{0}(\xi_{s})  =  
\boldsymbol{w}_{s}  \,  f'_0 (\xi_{s}) 
\text{ and } 
{\nabla}_{\mathbf r} \, f_{0}(\xi_{s}) 
=  - \left[ {\nabla}_{\mathbf r} \mu 
+ \left (\xi_{s} - {\mu} \right ) 
 \frac{  {\nabla}_{\mathbf r} T }{ T } \right]   f'_0 (\xi_{s})\, ,
\end{align}
where the``prime'' superscript is used to indicate partial-differentiation with respect to the variable shown within the brackets [for example, $ f_0^\prime (u) \equiv \partial_u f_0 (u)$]. Henceforth, we set $g_s= 1$, ignoring the degeneracy due to electron's spin.

In order to obtain a solution to the full BE, for small time-independent values of $\mathbf E$, $\nabla_{\mathbf r} T$, and $\nabla_{\mathbf r} \mu $, we assume a small deviation from the equilibrium distribution of the quasiparticles, such that
\begin{align}
\label{eqfdsoln}
f_s (\mathbf r,\mathbf k, t) \equiv  f_s (\mathbf r,\mathbf k)
	=  f_0 (\xi_{s}) +  \delta  f_s (\mathbf r,\mathbf k).
\end{align} 
Here, we have not included any explicit time-dependence in $\delta f_s (\mathbf r,\mathbf k) $, because the applied fields and gradients are static. Furthermore, we have suppressed showing explicitly the dependence of $f_0 $ on $\xi _s(\mathbf k)$, $\mu  (\mathbf r) $, and $T (\mathbf r)$.
We now parametrize the deviation as
$\delta  f _s (\mathbf r,\mathbf k)
= \sum \limits_{p=1}^\infty \epsilon^p \, f_{s}^{(p)}  $, where $\epsilon \in [0, 1 ]$ is the perturbative parameter. having the same order of smallness as the external perturbations $\mathbf E$, $\nabla_{\mathbf r} T$, and $\nabla_{\mathbf r} \mu $. Since $\epsilon$ is used solely for bookkeeping purpose (to track the order in the degree of smallness), we will set $ \epsilon =1$ at the end of the calculations, after we solve for $ \delta  f_s (\mathbf r,\mathbf k)$ recursively (i.e., order by order in increasing powers of $ \epsilon $).

\section{Useful integrals}
\label{secint}

In the main text, we have to deal with integrals of the form:
\begin{align}
\mathcal I = \int \frac{d^3 {\bf{k}} }{(2 \pi )^{3}} \, F ({\bf{k}} , \varepsilon_{s} ) \,
f_{0}^\prime ( \varepsilon_{s} )  \,,
\end{align}
where $ \varepsilon_{s} =  s \, v_0\, k + v_0 \,\eta \, k_z $. 
We switch to the spherical polar coordinates such that
\begin{align}
\label{eqpolar} 
k_x = \frac{ \tilde  \varepsilon \cos \phi \sin \theta} 
 {  s\, v_0}\,, \quad
 k_y = 
\frac{\tilde \varepsilon \sin \phi \sin \theta} { s\, v_0}\,, 
\quad k_{z} 
 = \frac{ \tilde  \varepsilon \cos  \theta} { s\, v_0} \,,
\end{align}
where $\tilde  \varepsilon \in [0, \infty )$, $\phi \in [0, 2 \pi )$, and $\theta \in [0, \pi ]$. The Jacobian of the transformation is $\mathcal{J} (\tilde  \varepsilon , \theta ) =   \frac{ \tilde  \varepsilon^2  \sin \theta}
{ s^3 \, v_0^3} $. This leads to
\begin{align}
\int_{- \infty}^{ \infty} d^3 \mathbf{k } \rightarrow   
\int_{0}^{ \infty} d  \tilde  \varepsilon  \int_{0}^{ 2 \,\pi} d \phi 
\int_{0}^{  \pi} d \theta \,   \mathcal{J} ( \tilde  \varepsilon , \theta ) 
\text{ and } 
 \varepsilon_{s}(\mathbf{k})  \rightarrow  
\varepsilon_{s}( \tilde  \varepsilon , \theta)   =  \tilde  \varepsilon \, 
\Lambda_s (\theta),  \text{ with } 
\Lambda_s (\theta) = 1 + \frac{\eta \cos{\theta} } {s} \,.
\end{align}

Since we have chosen the tilting direction with respect to the $z$-axis, the dispersion does not depend on $\phi$. Hence, we can perform the $\phi$-integration easily, after which $\mathcal I$ can be written in the following schematic form:
\begin{align}
\mathcal I &= \int_{0}^{ \infty} d \varepsilon   
\int_{0}^{  \pi} d \theta \,   \mathcal{I}_{1} ( \varepsilon, \theta) \, f^{\prime}_{0} (\varepsilon_s)\,.
\end{align}
Here, $ \mathcal{I}_{1} ( \varepsilon, \theta)$ is the function obtained after the $\phi $-integration.

In our computations, the $ \tilde \varepsilon $- and $\theta$-dependent parts of the integrand are decoupled. Hence, let us discuss some generic identities for these two kinds of integrals.
For the $ \tilde  \varepsilon$-integration, we encounter integrals of the forms $ \int_{0}^\infty d \tilde  \varepsilon \, 
{\tilde \varepsilon}^n \,  \, f^{\prime}_{0} (\varepsilon_s)$ and $\int_{0}^\infty d \tilde  \varepsilon \, \tilde  \varepsilon^n 
\,  \frac{\partial^{{ \lambda }+1} 
	\, f_{0} (\varepsilon_s)   } 
{ \partial \varepsilon_s^{{ \lambda }+1} }   $, where $\lbrace n, \lambda \rbrace \in \mathbb{Z}$.
Applying the Sommerfeld expansion \cite{mermin}, we get
\begin{align}  
\label{eqinte}
& \int_{0}^\infty d \tilde  \varepsilon \, \tilde \varepsilon^n \,  \, f^{\prime}_{0} (\varepsilon_s)    
= - \int_{0}^\infty d\tilde  \varepsilon \,\tilde  \varepsilon^n \,
\frac{\beta\, e^{\beta(\varepsilon_s- \mu)}}
{ \left [ 1 + e^{\beta \, \left (\varepsilon_s- \mu \right )} \right ] ^2}  
= - \frac{1}
{ \Lambda_s^{n+1} (\theta) }
\int_{0}^\infty  d {\tilde \varepsilon}   \, {\tilde \varepsilon}^n
\frac{\beta \, e^{\beta\, ( \tilde \varepsilon - \mu)}}
{(1 + e^{\beta \, ( \tilde \varepsilon- \mu)})^2} \
= -\, \frac{ \Upsilon_n ( \mu, T) }
{  \Lambda^{n+1}_s(\theta) } \,,\nn
\end{align}
where
\begin{align}
\label{equps}
\Upsilon_n ( \mu, T) =
\mu^n \left[
1 + \frac{\pi^2 \, T^2 \, n\, (n-1)}  {6   \, \mu^2} 
+ \frac{7 \, \pi^2 \, T^4 \,n \,(n-3) \,(n-2) \,(n-1) }
{360  \, \, \mu^4 } 
+ \order{\left(  {\frac{\mu}{T} } \right)^{-6}}
\right ],
\end{align}
which is valid in the regime $\beta \, \mu \gg 1$ (or $ \mu \gg T$ in the natural units). It is easy to show that \cite{ips-ruiz, ips-rsw-ph}, for higher-order derivatives, we have the relation
\begin{align}
\int_{0}^{ \infty} d \tilde  \varepsilon    \, \tilde   \varepsilon ^{n}  \,(-1)^{{ \lambda }+1}  \,
\frac{\partial^{{ \lambda }+1} 
	\, f_{0} (\tilde  \varepsilon  ) } { \partial  \tilde \varepsilon ^{{ \lambda }+1} }  
= \frac{n!}{(n-{ \lambda })!}  \, 
\frac { \Upsilon_{n-{ \lambda }} (\mu, T )} 
{{\Lambda_s^{n+1}(\theta)}}\,. 
\label{eqinte2}
\end{align}
For the $\theta $-integration, we use the identity
\begin{align}
& \int_{0}^\pi d\theta\,
\frac{ \left (\sin{\theta} \right )^{m} \left (\cos{\theta} \right)^n}
{( s +\eta  \cos{\theta})^l} 
= \int_{-1}^1 dt\,  
\frac{\left(1-t^2\right)^{\frac{m-1}{2}} \, t^n}{( s +\eta\,  t)^l}  \nn
& = \frac{\sqrt{\pi} \, \, \Gamma (\frac{m+1}{2})}  
{4 \, s^l} 
\Bigg [ 2 \,\left \lbrace  (-1)^n + 1 \right \rbrace 
\,  \Gamma \Big(\frac{n+1}{2} \Big) \,
\,   _3\tilde{F}_2 \Big(\frac{n+1}{2},\frac{l+1}{2},\frac{l}{2};
\frac{1}{2},\frac{1}{2} (m+n+2);\frac{\eta ^2}{s^2} \Big )    \nn
& \hspace{2.75 cm} 
+ \frac{\eta  \, l}{s}
\left \lbrace  (-1)^n - 1 \right \rbrace \,
\Gamma \Big(\frac{n}{2}+1\Big ) \, 
_3\tilde{F}_2 \Big(\frac{n+2}{2},\frac{l+1}{2},\frac{l+2}{2}; \frac{3}{2} ,\frac{m+n+3}{2} ;\frac{\eta ^2}{s^2} \Big)
\Bigg ]\,,
\end{align}
where $_{n_1}\tilde{F}_{n_2}\big( \{a_1,\ldots, a_{n_1} \};\{b_1,\ldots, b_{n_2} \}; X \big)$ is the regularized hypergeometric function \cite{hypergm}. 
 
 \section{Linear-in-$B$ parts of magnetoelectric conductivity}
\label{appsigma}

In this appendix, we outline the forms of the various components of the magnetoelectric conductivity tensor, involving the linear-in-$B$ parts, for the planar Hall set-up with $\mathbf E $ and $\mathbf B $ aligned parallel to each other (discussed in Sec.~\ref{sechall}).

\subsection{Intrinsic anomalous-Hall part}   
\label{appsigmaAH}

From the term proportional to $({\mathbf E}  \cross   \mathbf \Omega_s)$ in the integrand of Eq.~\eqref{eqcur},
we get the electric current density as
\begin{align}
 {\mathbf J}^{ {\text{AH}}}_s
& =   -\, e^2   \int
\frac{ d^3 \mathbf k}{(2\, \pi)^3 } \,
\left[   ({\mathbf E}  \cross   \mathbf \Omega_{s})   \right]
\,  f_0( \xi_s ) \,,
\end{align}
which gives the intrinsic anomalous-Hall term. Using $f_0(  \xi_s ) = f_0(\varepsilon_s  ) + \zeta_{s} \, f^{\prime}_0(\varepsilon_s ) + \order{ B^2}$, we get 
\begin{align}
(\sigma^{ {\text{AH}}}_s)_{a b} & = - \, e^2 \,\epsilon_{a b c}
\int \frac{  d^3 {\mathbf k} }  {(2 \,\pi )^3 } \, 
(\Omega_s )^c  
\left [ f_0  (\varepsilon_s )  
+ \zeta_s  \, f_0^\prime (\varepsilon_s ) + \order{ B^2}  \right ],
\end{align}
whose diagonal components are automatically zero because of the Levi-Civita symbol. Clearly, a nonzero OMM may generate $B$-dependent terms in $(\sigma^{ {\text{AH}}}_s)_{a b} $. 

For $\eta= 0 $, the contribution from the first term vanishes identically. For $\eta \neq 0$, although we get nonzero components from this first term, they are $B$-independent and, hence, do not contribute to $\varsigma^s_a$.
Let us consider the second term in the integrand, which is proportional to $\zeta_s  \, f_0^\prime (\varepsilon_s )$. First let us assume that $ \hat{\mathbf e}_E  = \hat{\mathbf x} $, which involves setting $ \zeta _{s} =  e \, \chi\, v_0 \, \mathcal{G}_s  \,{  k_ x\, B } /{k^2}  $. The corresponding contribution to
\begin{enumerate}

\item $(\sigma^{ {\text{AH}}}_s)_{y x} $ is
\begin{align}
t_{yx} \equiv - \, e^3 \, \chi\, v_0 \, \mathcal{G}_s  \, B \,
\epsilon_{y x z}
\int \frac{  d^3 {\mathbf k} }  {(2 \,\pi )^3 } \, 
(\Omega_s )^z  \, \frac{k_x} {k^2}  \, f_0^\prime (\varepsilon_s ) = 0\,;
\end{align}

\item $(\sigma^{ {\text{AH}}}_s)_{z x} $ is
\begin{align}
t_{zx} \equiv - \, e^3 \, \chi\, v_0 \, \mathcal{G}_s  \, B \,
\epsilon_{z x y}
\int \frac{  d^3 {\mathbf k} }  {(2 \,\pi )^3 } \, 
(\Omega_s )^y  \, \frac{k_x} {k^2}  \, f_0^\prime (\varepsilon_s ) = 0\,.
\end{align}

\end{enumerate}
Invoking the rotational symmetry of the dispersion about the $k_x k_y$-plane, we infer that, for $ \hat{\mathbf e}_E  = \hat{\mathbf y} $, we have $t_{xy} = t_{zy} =0 $.
Next, let us assume that $ \hat{\mathbf e}_E  = \hat{\mathbf z} $, which involves setting $ \zeta _{s} =  e \, \chi\, v_0 \, \mathcal{G}_s  \,{  k_z \, B } /{k^2}  $. The corresponding contribution to
\begin{enumerate}

\item $(\sigma^{ {\text{AH}}}_s)_{x z} $ is
\begin{align}
t_{xz} \equiv - \, e^3 \, \chi\, v_0 \, \mathcal{G}_s  \, B \,
\epsilon_{xzy}
\int \frac{  d^3 {\mathbf k} }  {(2 \,\pi )^3 } \, 
(\Omega_s )^y  \, \frac{k_z} {k^2}  \, f_0^\prime (\varepsilon_s ) = 0\,;
\end{align}

\item $(\sigma^{ {\text{AH}}}_s)_{y z} $ is
\begin{align}
t_{yz} \equiv - \, e^3 \, \chi\, v_0 \, \mathcal{G}_s  \, B \,
\epsilon_{yz x}
\int \frac{  d^3 {\mathbf k} }  {(2 \,\pi )^3 } \, 
(\Omega_s )^x  \, \frac{k_z} {k^2}  \, f_0^\prime (\varepsilon_s ) = 0\,.
\end{align}

\end{enumerate}

\subsection{Lorentz-force contribution}
\label{applor}

The leading-order contribution from the Lorentz-force part is obtained by picking up the
$n=1$ term in the expression for $\tilde g_s$ [shown in Eq.~\eqref{eqbolinter}], i.e., by using 
\begin{align}
 \delta f^{\chi}_{s} (\mathbf{k})
=  e^2 \, \tau^2  \, \mathcal{D}_{s} 
\, f_0^\prime (\xi_s) \,\hat{L}
\left [  \mathcal{D}_{s}   \left ( \boldsymbol{w}_{s} 
+ \boldsymbol{W}_{s} \right ) \cdot \mathbf{E}   \right ] .
\end{align}
This leads to the conductivity \cite{ips-rsw-ph}
\begin{align}
\label{eqsigLF}
& (\sigma_s^{\rm LF})_{ a b } = - \,\epsilon_{b d c} \, e^3 \, 
\tau ^2 \, s^3 \, v_0^3 
\int \frac{ d^3 \mathbf{k}} {(2 \, \pi)^3} 
\, \frac{ \mathcal{D}_s^2} {  \varepsilon_s^5} \,
 \left[ (w_{s})_a + (W_{s})_a \right ]   \,
  \left ( \varepsilon_s ^2 - \lambda_s \right )^2  \, \frac{B_c \,
  k_d \, f^\prime_0 (\xi_{s}) } {k} \,,\nn &
\lambda_s = 2 \, \chi \, e\, s \,  
{\mathcal{G}}_s \, v_0^2 \sum_{a'} \varrho_{a'} {B}_{a'}\, ,
\quad  \boldsymbol{\varrho} =  \cos{\phi} \sin \theta  \, {\boldsymbol{\hat x}}
 + \sin{\phi} \sin \theta \, {\boldsymbol{\hat y}} + \cos \theta  \, {\boldsymbol{\hat z}}  \,,
\end{align}   
where $\theta$ and $\phi $ refer to the polar and azimuthal angles used in the spherical-polar-coordinate transformation in Eq.~\eqref{eqpolar}. Using the same arguments as for the intrinsic anomalous-Hall part, we find that all the linear-in-$B$ components of
$ (\sigma_s^{\rm LF})_{ a b } $ vanish.

\subsection{Non-anomalous-Hall contribution}   
\label{appbcomm}

The non-anomalous-Hall part of the current is given by [cf. Eq.~\eqref{eqcur}]
\begin{align}
{\mathbf{\bar J}}_s
& =   -\, e^2\,\tau   \int
\frac{ d^3 \mathbf k}{(2\, \pi)^3 } \,
\left (  \boldsymbol{w}_s
+ \boldsymbol{W}_{s}  \right )
\mathcal{D}_{s}\left [     \left ( \boldsymbol{w}_{s} 
+ \boldsymbol{W}_{s} \right ) \cdot \mathbf{E}   \right ] f_0^\prime ( \xi_s) \,,
\end{align}
leading to
\begin{align}
\left(\bar \sigma_{s} \right)_{a b} 
= - \,e^2 \, \tau  
\int \frac{ d^3 \mathbf k}{(2\, \pi)^3 } \, \mathcal{D}_{s} 
\left[  (w_{s})_a \, + (W_{s})_a \right ]
\left [ (w_{s})_b \, + (W_{s})_b \right] \, f^\prime_0 (\xi_{s})  \,.
\end{align}
We want to compute here the linear-in-$B$ part of ${\bar \sigma}_{s}  $, after dividing it up as
$
  \sigma^{ \text{BC}}_{s} + \sigma^{ m}_{s} \,.$
Here, $\sigma^{ \text{BC}}_s $ arises solely due to the effect of the BC and survives when OMM is set to zero, while
$ \sigma^m_s $ is the one which goes to zero if OMM is ignored.

\subsubsection{BC-only part (no OMM)}

The BC-only part is given by 
\begin{align}
(\sigma^{ \text{BC}}_{s})_{a b} &= 
- \, \frac{e^3 \, \tau }{(2 \, \pi)^3} 
\int d^3 \mathbf{k} \,  \mathcal{M}_{a b}   \,  f^\prime_0  (\xi_{s} )\,, 
\quad
\mathcal{M}_{a b}  = \left[- \frac{(v_{ s })_a \, (v_{ s })_b }
{2}  \, 
\left( \mathbf \Omega_s \cdot \mathbf B \right )
+   (v_{ s })_a  
\left( \mathbf v_s \cdot \mathbf \Omega_s \right ) B_b \right ] 
+ a \leftrightarrow b  \,.
\end{align}

\subsubsection{Part with the integrand proportional to nonzero powers of OMM}

In order to actually carry out the integration for this part, it is convenient to express $\mathbf u_s $ as 
\begin{align}
(u_s)_{a} &= \sum_b (\mathcal{U}_{s})_{a b} \, B_{b} \,, \quad
\mathcal{U}_{s} = \begin{bmatrix}
	\Delta_{11} &   \Delta_{12} &  \Delta_{13}  \\
	\Delta_{21} &   \Delta_{22} &   \Delta_{23}  \\
	\Delta_{31} &   \Delta_{32} &  \Delta_{33}
\end{bmatrix} , 
\end{align}
where
\begin{align}
\Delta_{11}( \tilde \varepsilon, \theta, \phi) &= - \,
\chi \, e \, \mathcal{G}_s \, s^2 \, v_0^3 \;
\frac{  \sin^2 \theta  \cos (2 \phi )
	-\cos^2 \theta  } { \tilde \varepsilon^2}\,,\quad
\Delta_{22} ( \tilde \varepsilon, \theta, \phi) = 
\Delta_{11}( \tilde \varepsilon, \theta, \frac{\pi}{2} - \phi) \,,\nonumber \\
\Delta_{33} ( \tilde \varepsilon, \theta, \phi) &= -\,
\Delta_{11}( \tilde \varepsilon, \theta, 0)  \,,\quad
\Delta_{12} ( \tilde \varepsilon, \theta, \phi) 
= \Delta_{21} ( \tilde \varepsilon, \theta, \phi) 
= -\, \chi \, e \, \mathcal{G}_s \, s^2 \, v_0^3 \;   
\frac{\sin^2 \theta  \,\sin (2 \phi )} { \tilde \varepsilon^2} \,,\nonumber \\
\Delta_{13} ( \tilde \varepsilon, \theta, \phi) &
= \Delta_{31} ( \tilde \varepsilon, \theta, \phi) 
= \frac{\Delta_{12} } {\tan{\theta} \sin{\phi}} \,,\quad
\Delta_{23} ( \tilde \varepsilon, \theta, \phi) = 
\Delta_{32} ( \tilde \varepsilon, \theta, \phi)
= \frac{\Delta_{12} }{\tan{\theta} \cos{\phi}} \,.
\end{align} 
We can now express the relevant part of conductivity as 
\begin{align}
& (\sigma^{ m}_{s})_{a b}  = - \, e^2 \, \tau 
\int \frac{d^3 \mathbf{k}}{(2 \, \pi)^3}  
\left[ \mathcal{S}_{a b}  \, f^{\prime}_0 (\xi_{s} )  
+    \mathcal{P}_{a b}     \, f^{\prime \prime}_0 (\xi_{s} ) \right] , \nn
& \mathcal{S}_{a b}  =   \left [
(v_s)_{a} 
\sum_c (\mathcal{U}_{s})_{b c} \, B^c \right ]   + a \leftrightarrow b \,,\quad
\mathcal{P}_{a b} = \left [ - \frac{ (v_s)_a \, (v_{ s })_b} {2} 
\left( \mathbf m_s \cdot \mathbf B \right)
 \right ] + a \leftrightarrow b \,.
\end{align}

\subsection{Final expressions}
\label{secfinvarsig}

Expressing the linear-in-$B$ part of the current and the conductivity as
$ J^{{\rm lin}, s}_{a} = \left( \sigma_s^{\rm lin} \right)_{a  b} \, E^b  $
and $ \left( \sigma_s^{\rm lin} \right)_{a  b} $, respectively,
let us define a third-rank tensor
\begin{align}
\left( \varsigma_s^{\rm lin} \right )^c_{\; \;a b} 
= \frac{\partial \left( \sigma_s^{\rm lin} \right)_{a  b} }
{\partial B_c} \,.
\end{align}
For our scenario with $\mathbf E \parallel \mathbf B $, we will be dealing with the vector defined as
\begin{align} 
\varsigma^s_{a}=
\sum_{b, c} \left( \varsigma_s^{\rm lin} \right )^c_{\; \;a b}  \,\delta^b_{\;\; c} 
= \sum_{b} \left( \varsigma_s^{\rm lin} \right )^b_{\; \;a b} \,.
\end{align}
Gathering all the ingredients from the preceding subsections, we divide up $\varsigma^s_{a}$
as
\begin{align}
\varsigma^s_{a} = \varsigma_{a}^{(s,1)} + \varsigma_{a}^{( s, 2)}
+ \varsigma_{a}^{(s, 3)} + \varsigma_{a}^{(s, 4)} +\varsigma_{a}^{(s, 5)}  \,,
\end{align}
where
\begin{align}
\varsigma_{a}^{( s, 1)} & = e^3 \, \tau \sum_b
 \int \frac{ d^3 \mathbf{k}}
 {(2 \, \pi)^3}
  \left( v_s \right)_a \left( v_s \right)^b
 \left(   \Omega_s\right)_b   f^{\prime}_{0} (\varepsilon_s) 
= e^3 \, \tau 
 \int \frac{ d^3 \mathbf{k}}
 {(2 \, \pi)^3}
  \left( v_s \right)_a \left (\boldsymbol{v}_{s} \cdot \mathbf{\Omega}_{s} \right )   f^{\prime}_{0} (\varepsilon_s)  \,,\nn
\varsigma_{a}^{(s, 2)} & = -\, e^3 \, \tau 
 \sum_b \left( 1 +\delta_{a b} \right)
 \int \frac {d^3 \mathbf{k}}  {(2 \, \pi)^3}
  \left( v_s \right)_a 
 \left (\boldsymbol{v}_{s} \cdot \mathbf{\Omega}_{s} \right ) 
  f^{\prime}_{0} (\varepsilon_s) = - 4 \, \varsigma_{a}^{( s, 1)} \,,\nn
\varsigma_{a}^{(s, 3)}  & = e^2 \, \tau
\sum_b
\int \frac{d^3 \mathbf{k} }
{(2 \, \pi)^3} \left( v_s \right)_a 
\left [ \partial_{k_b}  \left( m_s \right)_b \right ]
f^{\prime}_{0} (\varepsilon_s)
= \frac{e^3\, \tau \, v_0 \, \mathcal{G}_s} {s}
\int \frac{d^3 \mathbf{k} }
{(2 \, \pi)^3} \left( v_s \right)_a 
\left [\nabla_{\mathbf k}  \cdot 
\left( k\, \mathbf \Omega_s \right) \right ]
f^{\prime}_{0} (\varepsilon_s) \; 
[\text{since } 	
\mathbf{m}_s ( \mathbf k) = \frac{e \, v_0 \, \mathcal{G}_s \, k } {s} 
	\, {\mathbf \Omega} _{s}( \mathbf k)]
\,,\quad
\nn \varsigma_{a}^{(s, 4)}  &= e^2 \,\tau 
\sum_b
\int \frac{d^3 \mathbf{k} }
{(2 \, \pi)^3} \left( v_s \right)^b 
\left [ \partial_{k^a } \left( m_s \right)_b \right ]
 f^{\prime}_{0} (\varepsilon_s) 
 =  \frac{e^3\, \tau \, v_0 \, \mathcal{G}_s} {s} 
\int \frac{d^3 \mathbf{k} }
{(2 \, \pi)^3} \left [  \boldsymbol v_s \cdot  
  \partial_{k_a } (k \, \mathbf \Omega_s ) \right ]
 f^{\prime}_{0} (\varepsilon_s) \,,\nn
\varsigma_{a}^{(s, 5)}  & = e^2 \, \tau \sum_b
\int \frac{d^3 \mathbf{k} } {(2 \, \pi)^3}
\left( v_s \right)_a \left( v_s \right)^b 
\left( m_s \right)_b f^{\prime \prime}_{0} (\varepsilon_s)
=
 \frac{e^3\, \tau \, v_0 \, \mathcal{G}_s} {s}
 \int \frac{d^3 \mathbf{k} } 
 {(2 \, \pi)^3}
\left( v_s \right)_a k
\left( \boldsymbol v_s \cdot 
\mathbf  \Omega_s \right) f^{\prime \prime}_{0} (\varepsilon_s)\,.
\end{align}  
While $\varsigma_{a}^{( s, 1)}$ and $\varsigma_{a}^{( s, 2)}$ represent the parts arising purely from the BC (i.e., with no contribution from the OMM), $\varsigma_{a}^{( s, 3)}$, $\varsigma_{a}^{( s, 4)}$, and $\varsigma_{a}^{( s, 5)}$ are the terms which go to zero if OMM is neglected.

Below, we elucidate the final expressions:
\begin{enumerate}

\item Noting that $ \boldsymbol v_s \cdot  \mathbf \Omega_s  = 
- \,\frac{\chi \, v_0 \,s} {k^2} \left( s + \eta \cos \theta \right) $, and using Eq.~\eqref{eqinte},
we obtain
\begin{align}
\varsigma_{a}^{( s, 1)} & = -\, \delta_{z a} \,\chi \, v_0 \, s \,
e^3 \, \tau  \int \frac{ d^3 \mathbf{k}}
 {(2 \, \pi)^3}
  \left( \eta + s \cos \theta \right) \frac{s +\eta \cos \theta} {k^2}   
\,   f^{\prime}_{0} (\varepsilon_s)
=  -\, \frac{ \delta_{z a} \, \eta\, v_0 \,  e^3 \, \tau 
\, \mathcal{C}_s }
{ 4 \, \pi^2   }  \,.
\end{align}  

\item Noting that $ \nabla_{\mathbf k}  \cdot 
\left( k\, \mathbf \Omega_s \right)  = -\,\chi \, s / k^2 ,$ and using Eq.~\eqref{eqinte}, we obtain
\begin{align}
\varsigma_{a}^{(s, 3)}  & 
= - \, \frac{\delta_{z a} \, \chi \,
e^3\, \tau \, v_0^2 \, \mathcal{G}_s} {(2 \, \pi)^3}
\int d^3 \mathbf{k} \,
\frac{ \eta + s \cos \theta  } {k^2} \,
f^{\prime}_{0} (\varepsilon_s)
\nn & = - \, \frac{\delta_{z a} \, \chi \,
e^3\, \tau \, v_0^2 \, \mathcal{G}_s} {(2 \, \pi)^2 }
\int d\theta \int d\tilde \varepsilon \,
 \frac{ \tilde \varepsilon^2 \, \sin \theta}{s^3 \, v_0^3}\,
\frac{ \eta + s \cos \theta  } {k^2} \,
\frac{s \, f_0^\prime (\tilde \varepsilon )} {s +\eta \cos \theta } 
\nn & = -\, \frac{
\delta_{z a} \, \chi \, v_0 \, e^3\, \tau  \, \mathcal{G}_s
} { 2\, \pi^2 } \,
\frac {\left (s^2 - \eta^2 \right)
\tanh^{-1}\big (\frac {\eta} {s} \big) - \eta  \, s } 
{\eta^2} \text{ for } |\eta| < |s| \,.
\end{align}  

\item Noting that $ \boldsymbol v_s \cdot  
  \partial_{k_a } (k \, \mathbf \Omega_s )  = - \,\chi \, s
  \left( v_s\right)_a / k^2 ,$ we obtain
\begin{align}
\varsigma_{a}^{(s, 4)}  & 
=  -\, \frac{ \delta_{za} \,\chi \,s\,
e^3\, \tau \, v_0 \, \mathcal{G}_s} {s} 
\int \frac{d^3 \mathbf{k} }
{(2 \, \pi)^3} \,\frac{\left( v_s \right)_z} {k^2}
 f^{\prime}_{0} (\varepsilon_s) 
 =  \varsigma_{a}^{(s, 3)}\,. 
 \end{align}  

\item
Noting that $ \boldsymbol v_s \cdot  \mathbf \Omega_s  = 
- \,\frac{\chi \, v_0 \,s} {k^2} \left( s + \eta \cos \theta \right) $, we obtain
\begin{align}
\varsigma_{a}^{(s, 5)}  & =
- \,\delta_{z a} \, \chi \, e^3\, \tau \, v_0^3 \, \mathcal{G}_s
 \int \frac{d^3 \mathbf{k} } 
 {(2 \, \pi)^3}
\left( \eta + s \cos \theta \right)
\frac{ s + \eta \cos \theta } {k}
\, f^{\prime \prime}_{0} (\varepsilon_s)
 = -\, \varsigma_{a}^{(s, 3)} \,. 
\end{align}

\end{enumerate}


\bibliography{ref_rsw}

\end{document}